\documentclass[fleqn,usenatbib]{mnras}

\usepackage{newtxtext,newtxmath}

\usepackage{booktabs}

\usepackage[T1]{fontenc}
\usepackage{amssymb}
\DeclareRobustCommand{\VAN}[3]{#2}
\let\VANthebibliography\thebibliography
\def\thebibliography{\DeclareRobustCommand{\VAN}[3]{##3}\VANthebibliography}

\usepackage{graphicx}	
\usepackage{amsmath}	
\usepackage{subcaption}
\usepackage{color}

\newcommand{\orcid}[1]{\href{https://orcid.org/#1}{\includegraphics[width=0.7em]{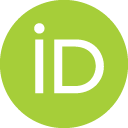}}}

\def\bigstrut{\vrule width0pt height0.5truecm}

\def\mpcoh{{\,h^{-1}\,\rm Mpc}}
\def\kmsmpc{\,{\rm km\,s}^{-1}{\rm Mpc}^{-1}}
\def\planck{{\it Planck\/}}
\def\citejap#1{\citeauthor{#1}\ \citeyear{#1}}
\def\gs{\mathrel{\lower0.6ex\hbox{$\buildrel {\textstyle >}
 \over {\scriptstyle \sim}$}}}
\def\ls{\mathrel{\lower0.6ex\hbox{$\buildrel {\textstyle <}
 \over {\scriptstyle \sim}$}}}
\def\vec#1{{\bf #1}}

\title[Clustering in the DESI Legacy Survey and its imprint on the CMB]{Galaxy clustering in the DESI Legacy Survey and its imprint on the CMB}

\author[Qianjun Hang et al.]
{Qianjun Hang \thanks{E-mail: qhang@roe.ac.uk},
Shadab Alam \orcid{0000-0002-3757-6359},
John A. Peacock
and Yan-Chuan Cai
\\
\bigstrut
Institute for Astronomy,
University of Edinburgh,
Royal Observatory,
Blackford Hill,
Edinburgh EH9 3HJ, UK\\
}

\pubyear{2020}

\begin{document}
\label{firstpage}
\pagerange{\pageref{firstpage}--\pageref{lastpage}}
\maketitle

\begin{abstract}
We use data from the DESI Legacy Survey imaging to probe the galaxy density field in tomographic slices covering the redshift range $0<z<0.8$. After careful consideration of completeness corrections and galactic cuts, we obtain a sample of $4.9\times 10^7$ galaxies covering 17\,739 deg$^2$. We derive photometric redshifts with precision
$\sigma_z/(1+z)=0.012 - 0.015$, and compare with alternative estimates. Cross-correlation of the tomographic galaxy maps with \planck\ maps of CMB temperature and lensing convergence probe the growth of structure since $z=0.8$. The signals are compared with a fiducial \planck\ $\Lambda$CDM model, and require an overall scaling in amplitude of $A_\kappa=0.901\pm 0.026$ for the lensing cross-correlation and $A_{\rm ISW} = 0.984 \pm 0.349$ for the temperature cross-correlation, interpreted as the Integrated Sachs-Wolfe effect. The ISW amplitude is consistent with the fiducial $\Lambda$CDM prediction, but lies significantly below the prediction of the AvERA model of R\'acz et al. (2017), which has been proposed as an alternative explanation for cosmic acceleration. 
Within $\Lambda$CDM, our low amplitude for the lensing cross-correlation requires a reduction either in fluctuation normalization or in matter density compared to the
\planck\ results, so that $\Omega_m^{0.78}\sigma_8=0.297\pm 0.009$. In combination with the total amplitude of CMB lensing, this favours a shift mainly in density: $\Omega_m=0.274\pm0.024$. We discuss the consistency of this figure with alternative evidence. A conservative compromise between lensing and primary CMB constraints would require $\Omega_m=0.296\pm0.006$, where the 95\% confidence regions of both probes overlap.
\end{abstract}




\section{Introduction}

The temperature fluctuations in the Cosmic Microwave Background (CMB) offer rich information about conditions in the early Universe at $z\simeq 1080$  (e.g. \citejap{PlanckT2018}). Photons from the CMB also interact through gravity with the large-scale structures (LSS) that they traverse, inducing two major secondary effects: gravitational lensing and the Integrated Sachs-Wolfe effect (ISW).
CMB lensing consists of the deflection of CMB photons by foreground LSS; the strength of the effect is quantified via the lensing convergence $\kappa$, which provides a measure of the projected matter density fluctuations between last scattering and the present. When general relativity holds, $\kappa$ is related to the 3D gravitational potential $\Phi$ projected along the line of sight:
\begin{equation}
\kappa (\vec{\hat n}) = \frac{1}{c^2}\int _0^{r_{\rm LS}}\frac{r_{\rm LS}-r}{r_{\rm LS}\,r}\, \nabla^2 \Phi(\vec{\hat n}, r)\, dr,
\end{equation}
where $\vec{\hat n}$ is the position vector on the sky, $r=\int {c}/{H(z)}\,dz$ is the line-of-sight comoving  distance, $\nabla^2$ is the comoving Laplacian, and $r_{\rm LS}$ is the distance to the last scattering surface; a flat geometry is assumed.
Lensing distorts the background Gaussian CMB sky and creates non-Gaussian signatures,
whose detection allows the reconstruction of a map of the convergence \citep[e.g.][]{Hu2000,OkamotoHu2003,LewisChallinor2006}.
The Integrated Sachs-Wolfe effect (\citejap{SachsWolfe1967}; \citejap{ISW1990}) arises from the time-dependent gravitational potential $\dot\Phi$ causing the CMB temperature $T_{\rm CMB}$ to change. The induced temperature fluctuation $\Delta T(\hat n)$ is proportional to the line-of-sight integral of $\dot \Phi$:
\begin{equation}
    \frac{\Delta T(\vec{\hat n})}{T_{\rm CMB}} =\frac{2}{c^2}\int \dot\Phi(\vec{\hat n}, t)\, dt
    =\frac{2}{c^3}\int \dot\Phi(\vec{\hat n}, t)\,a\, dr,
\end{equation}
where $T_{\rm CMB}$ is the mean temperature of the CMB at $z=0$ and $a=1/(1+z)$ is the cosmic scale factor.
In the linear regime, the ISW signal is non-zero when the matter density of the Universe, $\Omega_{\rm m}$, deviates from unity. It is therefore sensitive to the linear growth of structure and dark energy. These two effects offer spatial and temporal information about gravitational potential fluctuations. They couple the CMB with foreground LSS, and can be detected via cross-correlation measurements. This will be the main focus of our present study.

Observations of CMB lensing have progressed hugely in recent years, with a full sky map of lensing convergence delivered by \planck\ 
\citep{PlanckLens2013,PlanckLens2015,PlanckLens2018}, and over 2100 deg$^2$ by ACTpol \citep{ACTlens2020}. Here, we correlate the \planck\ lensing and temperature maps with LSS traced by galaxies. A particular aim is to measure the ISW effect, which has the attraction of providing an independent probe of dark energy. However, ISW detections have been challenging because the signal is largest at low multipoles where substantial cosmic variance is unavoidable; the effect has therefore been detected with only modest significance \citep[e.g.][]{Ho2008, Giannantonio2008, PlanckISW}. The uncertainty of measurements for the redshift range beyond $z>0.5$ is particularly large, with some having null, or anti-correlations between LSS and the CMB \citep{Sawangwit2010}. This regime is of particular interest as it may provide key evidence for distinguishing $\Lambda$CDM from early dark energy or modified gravity models (e.g. \citejap{Renk2017}).

In this study, we will exploit galaxy samples from the newly released DESI Legacy imaging survey \citep{DeyLegacy2019}. This covers over 1/3 of the sky, with depth substantially greater than alternative large-area imaging such as SDSS or Pan-STARRS, and it is therefore invaluable for CMB cross-correlation studies. The photometric precision and wide wavelength coverage permits the construction of robust photometric redshifts, allowing us to perform cross-correlations between galaxy samples in multiple tomographic redshift bins and both the CMB lensing convergence map and the CMB temperature map. This provides more information by constraining the evolution of both the ISW and lensing signals, and in principle allows an empirical determination of the growth history of density fluctuations. Recent examples of this sort of work include \cite{ISW2018} for the ISW effect and \cite{Giannantonio2016, Singh2017,Doux2018, Peacock2018,2020JCAP...05..047K, Singh2020, Darwish2020} for CMB lensing. A particular goal for the present study is to extend the redshift range of the tomographic measurements from $z\ls0.5$ to $z\simeq 1$ using the Legacy Survey.

An initial motivation for the present work was suggestions that foreground perturbations of the temperature field have an amplitude that is substantially in excess of the standard $\Lambda$CDM ISW prediction (\citejap{Granett2008}; \citejap{Kovacs2019}). These claims were based on stacking the signal from specially chosen `superstructures', but we shall not consider this detailed approach in the present paper. Rather, our aim here is to establish the Legacy Survey as a tool for two-point CMB tomography and to present the basic cross-correlation results.  Much of our analysis is therefore devoted to the issue of photometric redshifts and their calibration. Having established to our satisfaction that this can be done, we use the galaxy autocorrelation results to eliminate galaxy bias, allowing the cross-correlation measurements to yield a direct probe of
foreground mass fluctuations. In the simplest $\Lambda$CDM cosmology, this yields constraints
on the $\Omega_m - \sigma_8$ plane, which turn out to be in some tension with the parameter values inferred from the primary CMB fluctuations.

This paper is structured as follows: Section~\ref{sec:theory} collects the necessary theory for predicting cross-correlation signals; Section~\ref{sec:legacysurvey} presents the Legacy Survey data, including the derivation of independent photometric redshift estimates; Section~\ref{sec:results} presents the observed harmonic-space correlation results; Section~\ref{sec:interp} discusses the implications of our results; Section~\ref{sec:summary} sums up.


\section{Cross-correlation theory }\label{sec:theory}

To measure the lensing and ISW signals associated with our galaxy sample, we will employ galaxy auto-clustering ($gg$) and the cross-correlations with CMB lensing ($g\kappa$) and with CMB temperature ($gT$). The theoretical predictions for these quantities in the $\Lambda$CDM model are as follows; we work in spherical harmonic space and
follow the notation in \citet{Peacock2018}.
The galaxy harmonic auto-correlation in the Limber--Kaiser approximation \citep{Limber1953, Kaiser1992} is given by
\begin{equation}
    \frac{\ell(\ell+1)}{2\pi}C^{gg}_{\ell}=\frac{\pi}{\ell}\,\int b^2\Delta^2 (k=\ell/r,z)\,p^2(z)\, \frac{H(z)}{c} r\,dz,
    \label{eq:clgg}
\end{equation}
where $b$ is galaxy bias, $\Delta^2(k,z)$ is the dimensionless matter power spectrum at redshift $z$ ($\Delta^2(k,z)=k^3P_{\delta\delta}(k,z)/2\pi^2$), and $p(z)$ is the redshift probability distribution function: $\int p(z)\, dz=1$.  We use {\sc Halofit} \citep{Halofit1, Halofit2} as implemented in CAMB \citep{CAMB} to model the non-linear matter power spectrum. 
Note that the corresponding equation, (7), in \cite{Peacock2018} is misprinted and lacks the factor $\ell(\ell+1)/2\pi$.
For the case of galaxy cross-correlations between different tomographic slices,  $p^2(z)\rightarrow p_1(z)p_2(z)$ in Eq.~\ref{eq:clgg}, where $p_1(z)$ and $p_2(z)$ are the redshift probability distributions of the two slices.
We would also in principle have different biases for the two slices, $b^2\rightarrow b_1 b_2$, although for tomographic slices with a single sample selection, the bias is purely a function of redshift. Note that the above theory ignores distortions from peculiar velocities and treats redshift as an exact radial coordinate. This would not be correct for shells with width $\sim 10\mpcoh$, but is negligible for the much thicker shells that we consider \citep{Nock2010}.

Similarly,
the theoretical galaxy-lensing convergence cross power spectrum is computed by
\begin{equation}
    \frac{\ell(\ell+1)}{2\pi}C^{g\kappa}_{\ell}=\frac{\pi}{\ell}\,\int b\Delta^2(k=\ell/r,z)\,p(z)K(r)\, r \,dz,
    \label{eq:clgk}
\end{equation}
where the lensing kernel is given by
\begin{equation}
K(r)=\frac{3H^2_0\Omega_m}{2c^2 a}\frac{r(r_{\rm LS}-r)}{r_{\rm LS}}.
\end{equation}
Finally, the galaxy ISW cross-correlation is given by \begin{equation}
    \frac{\ell(\ell+1)}{2\pi} C^{gT}_{\ell}=T_{\rm CMB}\,\frac{2\pi}{c^3}\int b\, \Delta^2_{\delta\dot{\Phi}}(k=\ell/r,z)/k\,p(z)\,a\,dz.
    \label{eq:clgt}
\end{equation}
$\Delta^2_{\delta\dot{\Phi}}(k,z)$ is the dimensionless matter-$\dot{\Phi}$ cross-power spectrum. In linear theory, it is given by
\begin{equation}
    \Delta^2_{\delta\dot{\Phi}}(k,z)=\frac{3H^2_0\Omega_m}{2k^2}\frac{H(z)\left(1-f_g(z)\right)}{a}\Delta^2(k, z),
    \label{eq:Pgt}
\end{equation}
where $\Delta^2(k, z)$ is again the dimensionless matter power spectrum, $D(z)$ is the linear growth factor, $a$ is the scale factor, and $f_g\equiv d\ln D/d\ln a \simeq\Omega^{0.55}_m(z)$ is the growth rate \cite[e.g.][]{Crittenden1996,Afshordi2004,Ho2008,Giannantonio2008,PlanckISW}. N-body simulations have suggested that small deviations from linear theory for \smash{$C^{gT}_{\ell}$} occur at $\ell  \gtrsim 50$, and Eq.~\ref{eq:clgt} becomes inaccurate \citep{Seljak1996,Cooray2002, Cai2009,Smith2009, Cai2010,Carbone2016}. This can be alleviated by using the full nonlinear matter power spectrum in Eq.~\ref{eq:Pgt}, while still assuming a linear coupling between the density and velocity fields \citep{Cai2010}. Thus {\sc Halofit} is used to model $\Delta^2(k, z)$ in Eq.~\ref{eq:Pgt}.

The above expressions for angular power spectra assume spatial
flatness. The Limber-Kaiser approximation is inaccurate at large scales \citep[e.g.][]{Hu2000,verde2000}. The agreement between the small angle approximation and the exact computation is about 15\% in power at $\ell=10$, but quickly improving to $<1$\% for $\ell>30$. 
In practice these deviations are statistically negligible, as we exclude the largest-scale modes with $\ell<10$ from our fitting, to allow for possible complications from combining several surveys in the sky (see section \ref{sec:legacysurvey}).  Because of cosmic variance, those very large-scale perturbations contain little statistical power. Note also that in principle the bias parameter may depend on scale, although it should tend to a constant in the linear limit as $k\rightarrow 0$; in practice we do allow for this scale dependence (see Section \ref{sec:results-clgg}).

In summary, Eqs~\ref{eq:clgg}, \ref{eq:clgk} \& \ref{eq:clgt} are the theoretical predictions to be compared with our measurements from observations. The combinations of them should in principle allow us to determine both cosmological parameters and nuisance parameters such as galaxy bias and uncertainties in the true redshift distribution of the galaxy samples. Most directly, we can determine the amplitudes of the CMB lensing and ISW signals associated with the late-time LSS galaxies, relative to the prediction of a fiducial cosmological model.
We take this to be the \planck\ 2018 cosmology, with $n_s=0.965$, $\sigma_8=0.811$, $\Omega_m=0.315$, $\Omega_{\rm b}=0.0493$, and $H_0=67.4$ \citep{Planck2018Param}. The cross-correlation measurements are made using the CMB temperature and lensing $\kappa$ maps and masks from the 2018 \planck\ data \citep{PlanckT2018,PlanckLens2018}, together with  maps of galaxy number densities from the DESI Legacy survey. We detail our galaxy sample in the next section.

\section{Legacy Survey data}\label{sec:legacysurvey}
\subsection{Selection}
\label{sec:selection}
The Legacy Imaging Survey \citep{DeyLegacy2019} is a combination of four different projects observed using three different instruments on three different telescopes: the Dark Energy Camera Legacy Survey (DECaLS) observed using the Dark Energy Camera \citep{2015AJ....150..150F} including data from DES \citep{2005astro.ph.10346T}, the Mayall $z$-band legacy Survey (MzLS) observed by the Mosaic3 camera \citep{2016SPIE.9908E..2CD} and the Beijing-Arizona Sky Survey (BASS) observed by the 90Prime camera \citep{2004SPIE.5492..787W}.
 Altogether covering an area of 17,739 deg$^2$, the survey is divided around ${\rm Dec}=33^\circ$ in J2000 coordinates, with the southern part included in DECaLS, and the northern part covered by BASS and MzLS. We use the publicly available Data Release 8\footnote{\url{http://legacysurvey.org/dr8/}}. The sources are processed and extracted using {\tt Tractor}\footnote{\url{https://github.com/dstndstn/tractor}} \citep{2016ascl.soft04008L}, with three optical bands, ($g$, $r$, $z$), and three WISE \citep{Wright2010} fluxes ($W_1$, $W_2$, $W_3$) available. Because of the shallower effective depth of the $W_2$ and $W_3$ bands, we only make use of $W_1$ (3.4$\,\mu$m).
We apply the following selections to the data:
\begin{enumerate}
\item PSF type objects are excluded. This step eliminates most stars and quasars.
\item \texttt{FLUX\_G|R|Z|W1}$>0$, i.e. fluxes for all four bands are detected. This is to ensure successful determination of photometric redshifts.
\item \texttt{MW\_TRANSMISSION\_G|R|Z|W1} are applied to the fluxes for Galactic extinction correction.
\item Magnitude cuts are applied with $g<24$, $r<22$, and $W_1<19.5$, where all magnitudes are computed by $m=22.5-2.5\log_{10}({\rm flux})$. The cuts in $g$ and $r$ are chosen as reasonable completeness limits from inspection of the number counts. The cut in $W_1$ further removes faint objects that are not well covered by the calibration sample. We experimented with imposing a brighter cut, and found that our main results were essentially unchanged if all limits were made 0.5 mag. brighter.
\end{enumerate}

In addition, Bitmasks\footnote{\url{http://legacysurvey.org/dr8/bitmasks}} are used to generate a survey completeness map, with ${\rm bits}=(0, 1, 5, 6, 7, 11, 12, 13)$ masked. These masks cover foreground contamination at pixel level, including bright stars, globular clusters, and incompleteness in the optical bands. To convert the mask to appropriate resolution for this work, we generate large number of randoms and bin them into a {\tt healpix} map \citep{Gorski2005} with $N_{\rm side}=128$, corresponding to a pixel area of $0.2$\,deg$^2$. The completeness map is obtained by the ratio of the number of randoms in each {\tt healpix} pixel with and without masking. The map is then upgraded to $N_{\rm side}=1024$ which is the resolution used for most of our analyses.

\subsection{Photometric redshifts}

One of the key pieces of information needed for interpreting observations of CMB-galaxy cross-correlations is the redshift distribution of the galaxy sample. A variety of methods have been developed over many years to estimate either the redshifts of individual galaxies or the redshift distribution of a galaxy sample  using broad band photometry (see \citejap{2020arXiv200103621S} for a review). Generally photo-$z$ estimates are either template based (e.g. {\sc LePhare} \citejap{1999MNRAS.310..540A}; {\sc BPZ} \citejap{2000ApJ...536..571B}; {\sc EAZY} \citejap{2008ApJ...686.1503B}) or data-driven methods (e.g. {\sc TPZ} \citejap{2013MNRAS.432.1483C}; {\sc SkyNet} \citejap{2014MNRAS.441.1741G}; {\sc GPz} \citejap{2016MNRAS.462..726A}; {\sc ANNZ2} \citejap{2016PASP..128j4502S}; {\sc METAPhoR} \citejap{2017MNRAS.465.1959C}; {\sc Delight}  \citejap{2017ApJ...838....5L}; {\sc CMN} \citejap{2018AJ....155....1G}; {\sc CHIPPR} \citejap{2020arXiv200712178M}) . There have been several attempts to compare the accuracy and precision of various photometric redshift methods \citep{2010A&A...523A..31H, 2015MNRAS.452.3710R, 2014MNRAS.445.1482S, 2016PhRvD..94d2005B} with no strong winner. 
Our approach is direct and empirical, based on using observed spectroscopy to assign a redshift to a given location in multi-colour space. In parallel with this work, a public catalogue of photometric redshifts for the Legacy Survey was made available by \cite{Zhou2020}; Z20 hereafter. Their approach differs somewhat from ours, being based on machine learning. The advantage of this is that we are able to look in detail at the sensitivity of our results to the properties of the photometric redshifts.

The following spectroscopic surveys are used for redshift calibration: GAMA \citep[DR2; ][]{2015MNRAS.452.2087L}, BOSS \citep[DR12; ][]{2015ApJS..219...12A}, eBOSS \citep[DR16; ][]{2020ApJS..249....3A}, VIPERS \citep[DR2; ][]{2018A&A...609A..84S}, and DEEP2 \citep{2013ApJS..208....5N}. In addition, we also include COSMOS \citep{2009ApJ...690.1236I} and DESY1A1 redMaGiC \citep{2018MNRAS.481.2427C} for their highly accurate photometric redshifts. The redshifts from the original calibration samples will be referred to as `spectroscopic' or `true' from this point onward, in order to make a distinction with the inferred photometric redshifts. The majority of these datasets overlap with DECaLS, and galaxies in the calibration data sets are matched with DECaLs objects based on their nearest neighbours using the python routine {\tt cKDTree} within a distance of 0.5$^\circ$. The GAMA sample has been rejection sampled to remove the dip in density around $z=0.23$; this is known to represent a rare LSS fluctuation, which we do not wish to imprint on our photo-$z$ estimates. For the COSMOS sample, only objects with \texttt{r\_MAG\_APER2} $<23$ are used to match with the DECALS sample. The whole calibration sample roughly covers the redshift range $0<z<1$.

\begin{figure}
	\centering
	\includegraphics[width=0.47\textwidth]{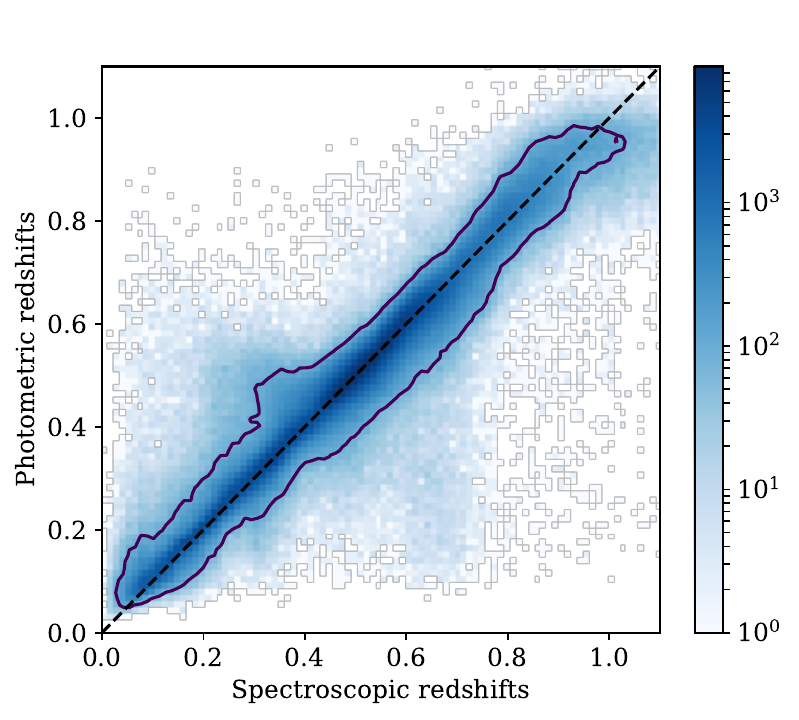}
	\caption{Photometric redshifts inferred from $g-r$, $r-z$, and $z-W_1$ colours, versus the spectroscopic redshifts for the calibration samples. The contour shows the 95\% interval. The colour bar indicates the number of galaxies in each pixel.}
	\label{fig:calibration_photoz_specz}
	\vspace{-0.2 cm}
\end{figure}

All calibration samples except DESY1A1 redMaGiC \citep{2018MNRAS.481.2427C} are binned in 3-dimensional grids of $g-r$, $r-z$, and $z-W_1$ with a pixel width of about 0.03. The range of the colours are: $-0.5<g-r<2.5$, $-2<r-z<3$, and $-2<z-W_1<4$. Pixels containing more than 5 objects from the calibration samples are assigned the mean redshift of these objects. The DES sample is processed in the same way to fill out pixels that are not calibrated in this initial pass. We then apply this calibration to the full Legacy Survey: objects that fall in pixels that lack a redshift calibration are excluded, thus selecting objects that occupy the same colour space as our calibration sample. The assigned photometric redshift is the mean redshift for the colour pixel, plus a random top-hat dither of $\pm0.005$ so that digitization artefacts are not apparent in the $N(z)$ distributions. Fig.~\ref{fig:calibration_photoz_specz} compares the inferred photometric redshifts with the true redshifts of the calibration sample. The general agreement is good, with $68\%$ of the sample having photometric redshifts within $\pm0.027$ of their spectroscopic redshifts. However, a small proportion of the objects with true redshifts $0.2<z<0.4$ are assigned photometric redshifts between $0.4<z<0.6$. The inferred redshifts are also underestimated beyond $z=0.9$, as usual: this estimation method means that $\langle z_{\rm spec}\rangle$ should be unbiased at given $z_{\rm phot}$, so that a bias in $\langle z_{\rm phot}\rangle$ at given $z_{\rm spec}$ is inevitable at the extremes of the distribution.

\begin{figure}
	\centering
	\includegraphics[width=0.47\textwidth]{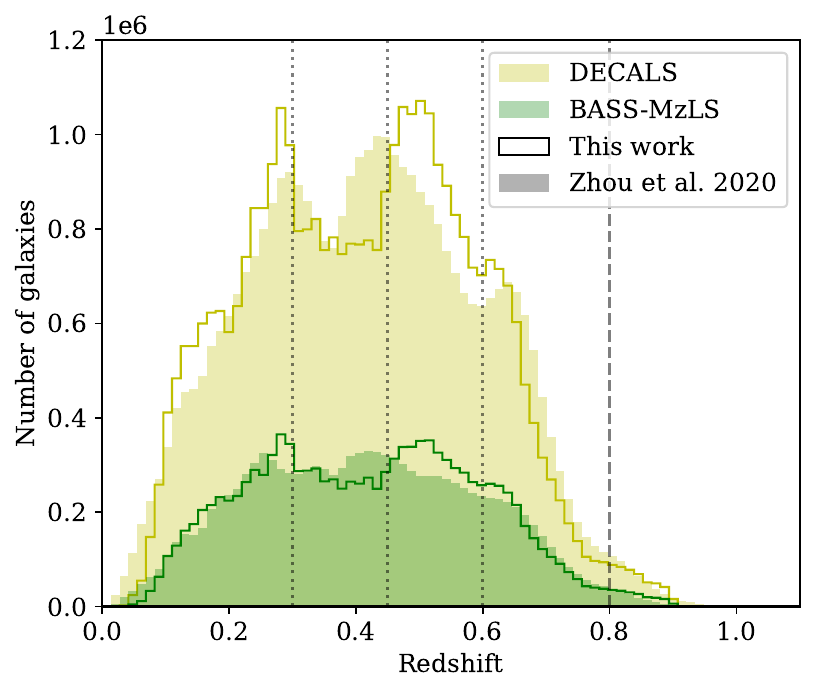}
	 \caption{Photometric redshift distribution of galaxies after selection, in the DECALS (yellow) and BASS-MzLS (green) regions respectively. We compare our photometric redshifts (shown as a solid line histogram) with the corresponding redshifts from \protect\cite{Zhou2020} (shown as a shaded histogram). Grey dotted lines show our four tomographic redshift bins in $0<z\leq0.8$.}
	\label{fig:inferz_photoz_comp}
\end{figure}

Photometric redshifts are assigned to $78.6\%$ of the selected Legacy Survey objects, yielding a primary sample of approximately 49 million galaxies (see Table~\ref{tab:redshift_error} for details). The lost $21.4\%$ objects will lead to higher shot noise, but this is a small price to pay for excluding objects where the photometric redshift cannot be reliably calibrated. The redshift distribution of our final sample is shown in Fig.~\ref{fig:inferz_photoz_comp}. We can compare this distribution with the corresponding $N(z)$ for the public Legacy Survey photometric redshifts made available by \cite{Zhou2020}; this is shown in Fig.~\ref{fig:inferz_photoz_comp}. The two distributions are generally in good consistency with each other; both distributions show some weak features, indicating that LSS in the calibrating samples has still propagated into the final photo-$z$s to some extent. With broad tomographic bins, we expect that such structure will be unimportant, but it will be helpful to compare the results from two rather different photo-$z$ catalogues. We divide our samples into four tomographic slices, illustrated by the grey dotted lines in Fig.~\ref{fig:inferz_photoz_comp}. The redshift ranges are: bin 0: $0<z\leq 0.3$; bin 1: $0.3<z\leq 0.45$; bin 2: $0.45<z\leq 0.6$; bin 3: $0.6<z\leq 0.8$. Our photo-$z$ data and accompanying software can be accessed at \url{https://gitlab.com/qianjunhang/desi-legacy-survey-cross-correlations}.

\subsubsection{Photometric redshift error distribution}

For the calibration sample, the distribution of $\delta z\equiv z_{\rm spec}-z_{\rm phot}$ as a function of $z_{\rm phot}$, can be well modelled by the modified Lorentzian function, 
\begin{equation}
   L(x)=\frac{N}{\left(1+((x-x_0)/\sigma)^2/2a\right)^a},
   \label{eq:lorentzian}
\end{equation}
where $x_0$, $\sigma$, and $a$ are parameters that control the mean. width, and fall-off of the distribution, and $N$ is the normalization such that $\int_{-\infty}^{+\infty}L(x)\,dx=1$. For each of the tomographic bins, we fit $\sigma$ and $a$, while $x_0$ is fixed to zero. These parameters are summarized in Table~\ref{tab:redshift_error}. The inferred true redshift distribution $p(z)$ is then estimated by convolution of the raw distribution with the Lorentzian function. \cite{2020arXiv200712795S} have recently proposed a similar approach to marginalizing over photo-$z$ errors while restricting themselves to the case of Gaussian fields with an ad-hoc mixing matrix.

\begin{table*}
	\centering
	\caption{Summary of the four tomographic redshift slices. The first row shows the number of galaxies in each redshift slice. The second row shows the effective volume of the redshift slice. The third and forth rows are parameters for the Lorentzian function (Eq.~\ref{eq:lorentzian}) fitted to redshift errors in each redshift bin derived from the calibration data sets; and the last two rows show the best-fit parameters derived empirically from the cross-correlations between the different tomographic bins (noting that $\sigma$ is not varied in this exercise). The best-fit parameters refer to our photo-$z$ data clipped with $|\Delta z|<0.05$.}
	\label{tab:redshift_error}
\begin{tabular}{ l|l|l|l|l }
 \hline
 Redshift bin & 0: $0<z\leq0.3$ & 1: $0.3<z\leq0.45$ & 2: $0.45<z\leq0.6$ & 3: $0.6<z\leq0.8$ \\ 
 \hline
 Number of galaxies & 14 363 105 & 11 554 242 & 13 468 310 & 7 232 579 \\
 Volume [($h^{-1}$Gpc)$^3$] & 1.047 & 2.084 & 3.431 & 20.37\\
 $\sigma^{\rm spec}$ & 0.0122 & 0.0151 & 0.0155 & 0.0265\\ 
 $a^{\rm spec}$ & 1.257 & 1.319 & 1.476 & 2.028\\ 
 $a^{\rm bf}$ & 1.257 & 1.104 & 1.476 & 2.019\\ 
 $x_0^{\rm bf}$ & \llap{$-$}0.0010 & 0.0076 & \llap{$-$}0.0024 & \llap{$-$}0.0042\\
 \hline
\end{tabular}
\end{table*}

\begin{figure}
	\centering
	\includegraphics[width=0.47\textwidth]{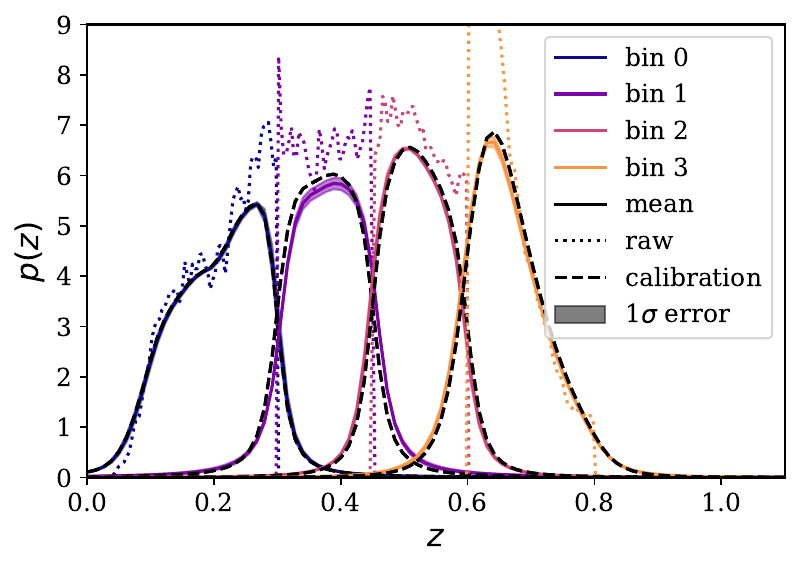}
	\caption{Redshift distribution function, normalized such that for each redshift bin $\int p(z)\,dz=1$. The dotted lines show the raw photometric redshift distribution with $|\Delta z|<0.05$, the solid lines show the mean distribution (see text for details) and their 1-$\sigma$ deviation using the 2-bias model, and the dashed lines show the distribution using parameters from spectroscopic calibration sample.}
	\label{fig:nz}
\end{figure}

However, galaxies fainter than the calibration sample may not follow this $\delta z$ distribution exactly. There is an irreducible scatter that arises because galaxy spectra are not universal in shape; but photometric measuring errors will increase the scatter for fainter objects. As shown below in Section~\ref{sec:results-clgg}, we are able to diagnose this using the galaxy cross-correlations between the different tomographic redshift slices. The width of the error distribution controls the degree of cross-correlation between the different tomographic slices, which is observed to be larger than predicted when using the directly calibrated $p(z)$ parameters from Table~\ref{tab:redshift_error}. The largest discrepancy occurs in the cross-correlation between redshift bin 1 and bin 2, which is almost double the predicted value. We therefore model the true error distribution in the photometric redshifts by allowing the tail $a$ of each distribution to spread, while fixing the width $\sigma$ to that determined by the spectroscopic sample. We also allow a change in the mean $x_0$ of each bin, while requiring the sum of the mean shifts in the four bins to be zero. This results in 7 systematic nuisance parameters to marginalize over. We take 10 samples in each dimension of the 7-parameter space with appropriate upper and lower bounds, and for each point in the grid, we compute the $\chi^2$ of the 10 galaxy auto- and cross-correlation between different redshift slices. The galaxy bias parameters in each case are fixed at the lowest-$\chi^2$ values from the auto-correlation (which we fit using the 2-bias model up to $\ell=500$). This is sufficient given the size of the error bar in the auto-correlations: the galaxy bias is very tightly constrained. Constraints on the cross-correlation amplitudes can then be marginalized over the photo-$z$ parameters, i.e., weighted by the likelihoods of each set of parameters. The mean and 1-$\sigma$ deviation of $p(z)$ weighted by the likelihoods of the $p(z)$ parameters are shown in Fig.~\ref{fig:nz}. The procedure is detailed in Section~\ref{sec:results-clgg}.

\subsection{Comparison with Zhou et al. (Z20)}

It is interesting to compare our redshift estimates with those of Z20: \cite{Zhou2020}. This is studied in some detail in Appendix~\ref{app:A}, but we summarise the main features here. Firstly note that this comparison is only possible for the 78.6\% of galaxies that lie in regions of multicolour space for which calibration data exist. Z20 give photometric redshifts for additional galaxies, and these are probably to be considered less reliable. Nevertheless, we can perform clustering analyses that use all the Z20 data, or just their redshifts for the same set of objects that we use, and this can give useful insight into the robustness of our conclusions. For the objects in common, the median redshift difference is $|\Delta z|=0.023$, and 68\% of objects agree in photometric redshift to within 0.038. The difference distribution has non-Gaussian tails, and we also therefore consider a `clipped' selection where we retain only objects where the two estimates agree to within $|\Delta z|<0.05$: this is about twice as large  as our photo-$z$ 1-$\sigma$ uncertainty, so the effect is to remove outlying objects in the tails of the error distribution. This removes a further 23.4\% of the sample, but should provide a cleaner selection in the sense that object are more likely to lie in their nominal tomographic bin. The cross-correlations between the different bins confirm that this strategy is successful.


\begin{figure*}
    \centering
    \begin{subfigure}[b]{0.49\textwidth}
        \includegraphics[width=\textwidth]{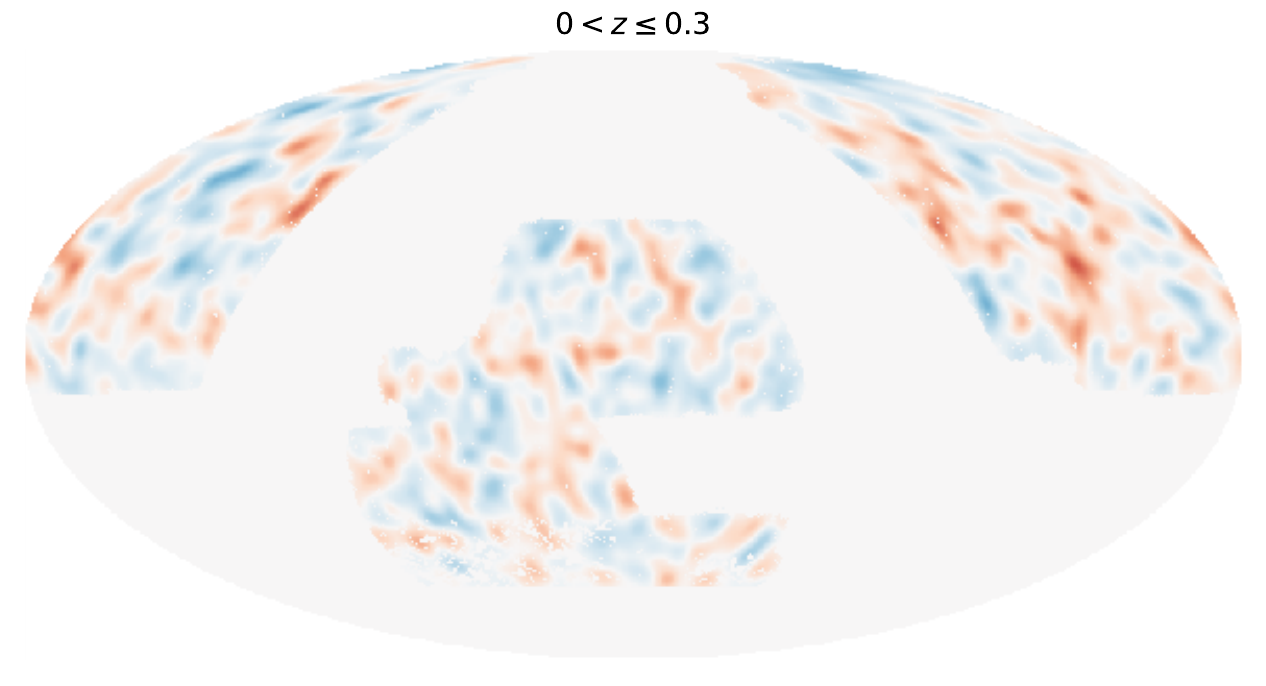}
    \end{subfigure}
    ~ 
    \begin{subfigure}[b]{0.49\textwidth}
        \includegraphics[width=\textwidth]{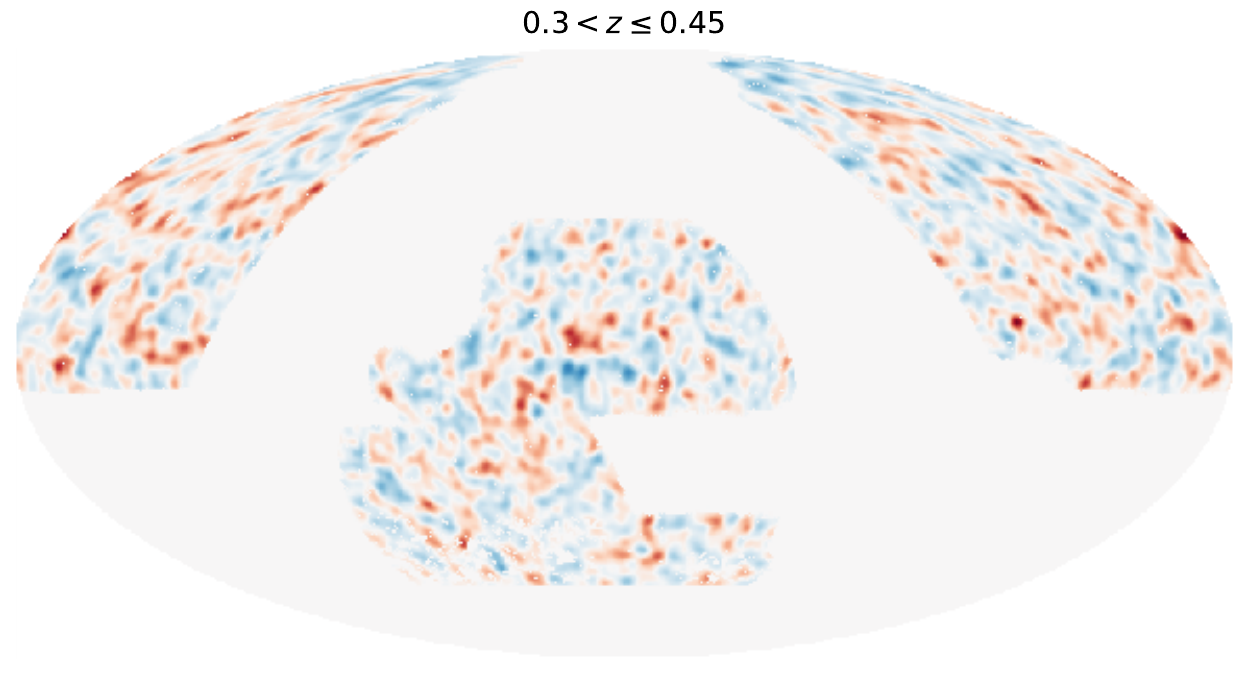}
    \end{subfigure}
    ~ 
    \begin{subfigure}[b]{0.49\textwidth}
        \includegraphics[width=\textwidth]{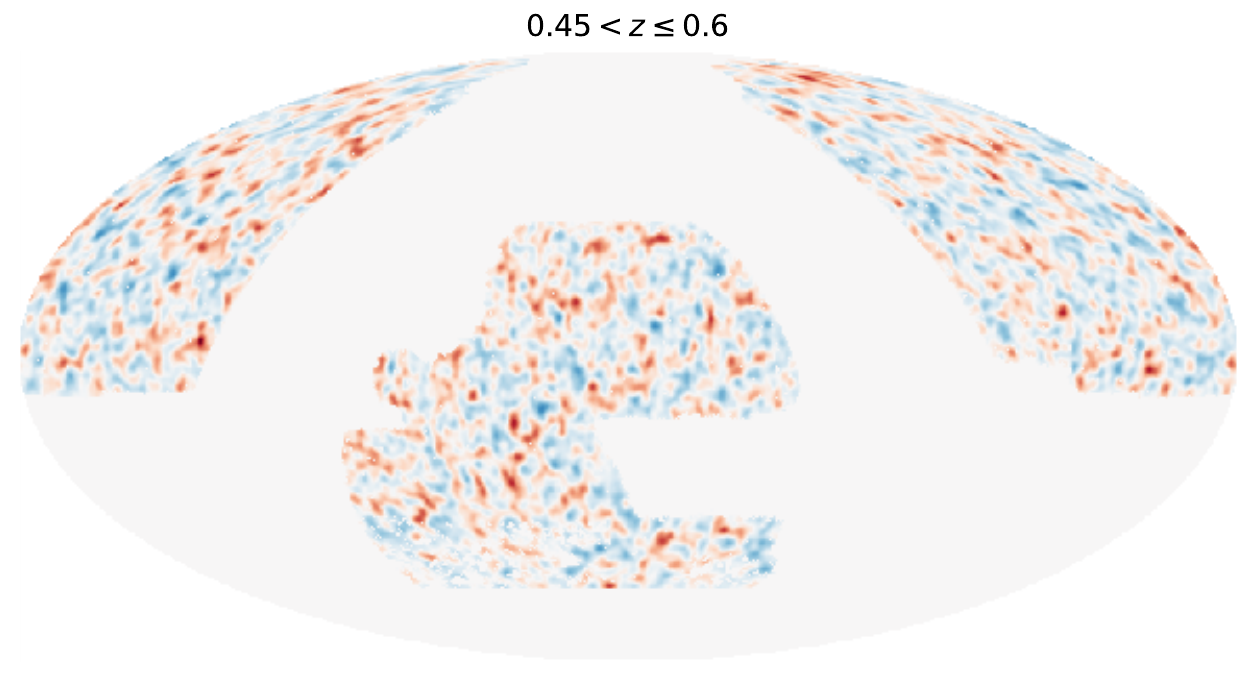}
    \end{subfigure}
    ~ 
    \begin{subfigure}[b]{0.49\textwidth}
        \includegraphics[width=\textwidth]{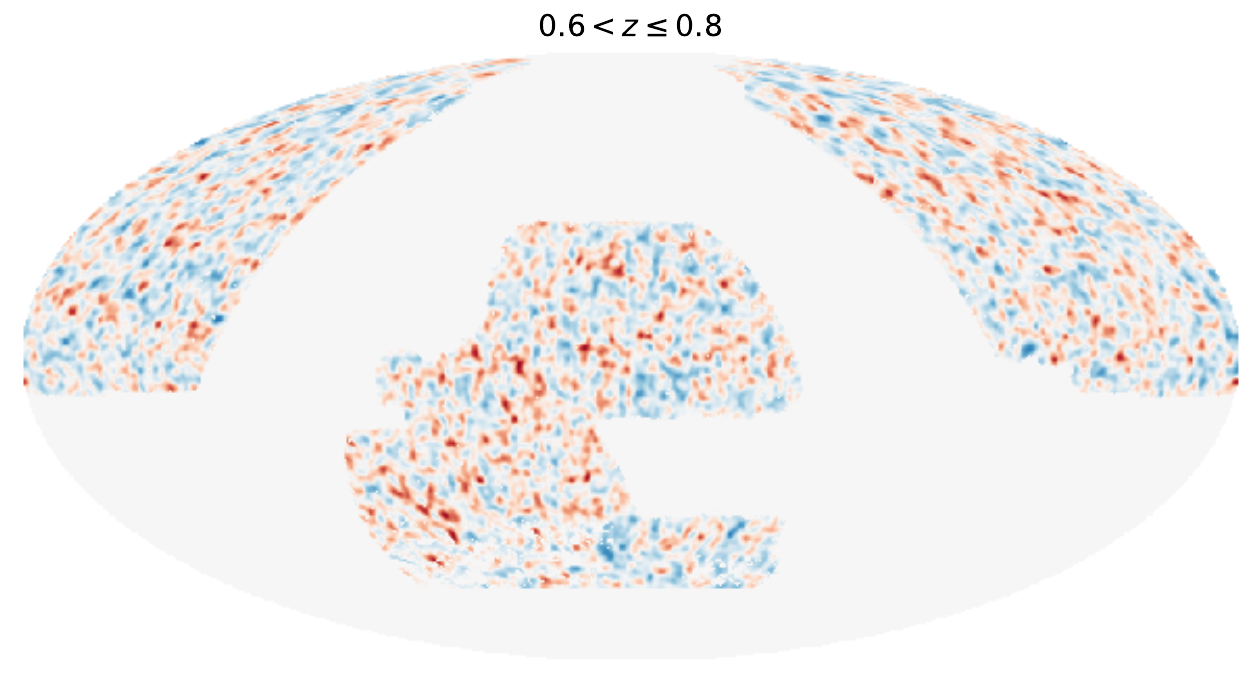}
    \end{subfigure}
    ~
    \begin{subfigure}[b]{\textwidth}
        \centering
        \includegraphics[width=0.5\textwidth]{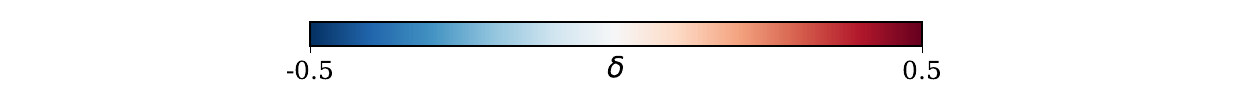}
    \end{subfigure}
    \caption{The density fluctuation maps for the four tomographic slices. For illustrative purpose only, they are smoothed by a Gaussian symmetric beam with comoving scale of $20\mpcoh$. These maps are made from galaxy maps via Eqn.~\ref{eq:delta}, and corrected by completeness and stellar density.}\label{fig:density_maps}
\end{figure*}

\subsection{Galaxy maps and systematic corrections}

Galaxies in each tomographic slice are binned in {\tt healpix} maps with $N_{\rm side}=1024$. The density fluctuation, $\delta$, in each pixel is then computed by
\begin{equation}
    \delta=\frac{n}{\bar{n}}-1,
    \label{eq:delta}
\end{equation}
where $n$ is number of galaxies in the pixel, and $\bar{n}$ is the mean number of galaxies per pixel. Due to the slight differences in the photometric passbands for DECam, BASS, and MzLS, the surface density of the tomographic slices varies slightly, between $2\%$ and $5\%$, in the north and south regions. For our purpose here, we compute $\delta$ for the north and south regions separately, and join the two regions at Dec $=33^\circ$.

The density maps are correlated with various systematics, including observational conditions, survey depth, stellar density, and Galactic extinction. Most foreground contamination is captured by the completeness map. In addition, we use the ALLWISE total density map as a proxy for stellar density. We find little correlation with the $E(B-V)$ extinction map. The following corrections are applied to the density map to remove possible systematics.

To obtain an unbiased mean density, we compute $\bar{n}$ using pixels with completeness $>0.95$ and stellar number 
$N_{\rm star}<8.52\times10^3$/deg$^2$,
about $70\%$ of the total unmasked pixels. The largest correlation with density comes from the completeness map. The galaxy count in each pixel is corrected by $n/w$, where $w$ is the completeness in each pixel. Regions with $w<0.86$ are masked, based on the binned one-dimensional relation between the completeness and mean density fluctuation in the bin, $\bar{\delta}$, such that the deviation of $\bar{\delta}$ from zero is smaller than 0.1. We also introduce a similar cut in stellar number at 
$N_{\rm star}<1.29\times10^4$/deg$^2$.
The residual binned one-dimensional correlation between $\log_{10}(N_{\rm star})$ and mean $\delta$ in the bins is below $5\%$ for all bins except for the highest redshift bin at the large stellar density end. We use 5th-order polynomials to fit for the residual correlation for each bin as a function of  $\log_{10}(N_{\rm star})$ and subtract the residual mean density from the raw $\delta$. The final corrected density maps are cross-correlated with the completeness map and stellar density map in each bin. The resultant correlation is consistent with zero for the $\ell$ range used in the analysis. The corrected density fluctuations in the four redshift slices are shown in Fig.~\ref{fig:density_maps}. For illustrative purpose, they are smoothed by a Gaussian symmetric beam with $\sigma=20\mpcoh$ in comoving distance. We note that the photometric variations and correlations with various foreground maps for our sample are relatively small. This is driven by the magnitude cuts used in our selection (see section ~\ref{sec:selection}). \cite{2020MNRAS.496.2262K} provides a more detailed analysis of photometric systematics for a variety of galaxy samples.

\begin{figure*}
	\centering
	\includegraphics[width=\textwidth]{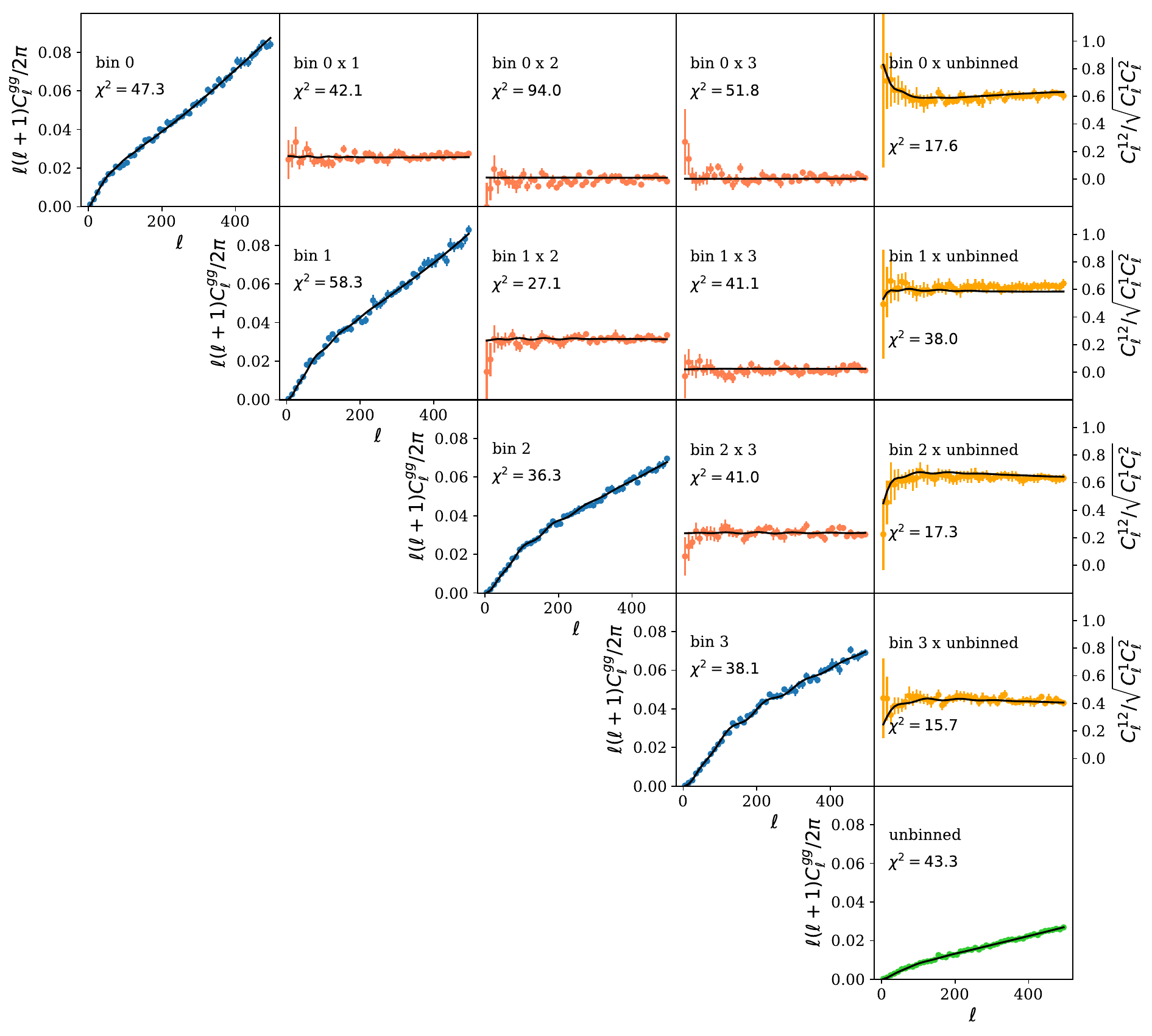}
	\caption{The galaxy auto-correlation $C_{\ell}^{gg}$ for each redshift slice (diagonal) and cross-correlation coefficients between different slices (off-diagonal). The last column shows the auto- and cross-correlations with the unbinned case, with shot noise subtracted. Data is presented in groups of 10 modes. The black solid line shows the theory with the best-fit $p(z)$ and redshift-dependent bias. The fitting of $p(z)$ is performed simultaneously for all the sub-sections except the last column  for modes in $10<\ell<500$, with a total DOF=$49\times10-7=483$ and the total $\chi^2=471$. The break-down of $\chi^2$ is in each case is shown on the top left corner of each sub-section.}
	\label{fig:clgg}
\end{figure*}

\section{Results}\label{sec:results}

We now compute the angular correlations between our various galaxy number density maps, and with the CMB $\kappa$ and temperature maps. We will be especially interested in comparing the amplitudes of the CMB lensing and ISW signals with the predictions of the fiducial $\Lambda$CDM model, from Eqs~\ref{eq:clgk} \& \ref{eq:clgt}. The procedure can be summarised as follows: \\
$\bullet$ Constrain linear galaxy biases with the galaxy auto- and cross-correlations from the four redshift bins:
\begin{equation}
C_{\ell}^{g_ig_j}=b_ib_jC_{\ell}^{\delta\delta}.
\end{equation}
Here, we allow the pdf of photo-$z$s to vary with nuisance parameters that will be marginalised over. \\
$\bullet$ Measure the amplitude of the lensing and ISW signals $A_{\kappa}$ and $A_{\rm ISW}$ defined as 
\begin{equation}
C_{\ell}^{g\kappa} = A_{\kappa}bC_{\ell}^{\delta\kappa}; \quad C_{\ell}^{gT} = A_{\rm ISW}bC_{\ell}^{\delta T},
\end{equation}
incorporating the constrained galaxy biases from the previous step.

The angular power $C_{\ell}$ is computed by converting a pixel map into its spherical harmonics $a_{\ell m}$ in {\tt healpy}. For a masked map, we use the simplest pseudo-power estimate $\hat{C}_{\ell}=C_{\ell}^{\rm masked}/f_{\rm sky}$. We have verified that inaccuracies in this estimate are unimportant for this large sky coverage, especially given that we exclude $\ell<10$ as further insurance against any residual large-scale systematics. We also impose an upper cutoff: throughout the analysis, we use modes in the range $10\le \ell<500$.  The $\ell>500$ modes give very noisy measurements for cross-correlations between LSS and CMB, and the $S/N$ for the amplitude of the cross-correlation signal has converged by this point. Linear bias is no longer a valid assumption beyond about $\ell=250$, and we make allowance for scale-dependent bias as described below. We use a {\tt healpix} resolution of $N_{\rm side}=1024$ for our analysis, and have tested that using finer maps would not alter the results. We correct for the pixel window function, although this is not a significant effect.

In the following analysis, we group every $M=10$ $\ell$-modes together such that
\begin{equation}
    \langle C_{\ell} \rangle_{\rm group}=\frac{1}{M}\sum_{\ell'}^{\ell'+M-1}C_{\ell'},\quad \ell'=M, 2M,...,
\end{equation}
and $\ell$ is the median value in each case. A simple error bar on each grouped data point can then be computed by
\begin{equation}
    \sigma_{\ell}=\frac{1}{f_{\rm sky}}\sqrt{\frac{\langle C^2_{\ell}\rangle - \langle C_{\ell}\rangle^2}{M-1}}.
\end{equation}
The $f_{\rm sky}$ factor takes care of correlations between $\ell$-modes due to the masked sky. We have verified using simulated lognormal density maps that this simple approximation indeed gives unbiased error estimates, leading to a diagonal covariance, $C={\rm diag}(\sigma_{\ell}^2)$. The $\chi^2$ of a theoretical model is defined as
\begin{equation}
    \chi^2=\mathbf{d}^TC^{-1}\mathbf{d},
    \label{eq:chi2}
\end{equation}
where the vector $\mathbf{d}$ has components $d_{\ell}=C_{\ell}^{\rm data}-C_{\ell}^{\rm th}$. 
The likelihood of a model parameter set $\vec x$ is given by
\begin{equation}
    \mathcal{L}(\vec x)=\frac{e^{-\chi^2(\vec x)/2}}{\int e^{-\chi^2(x)/2}\,d^nx},
\end{equation}
where as usual we will take the likelihood to give the posterior on the parameters, assuming uninformative uniform priors.

The theory vector \smash{$C_{\ell}^{\rm th}$} contains the predictions from Eqs~\ref{eq:clgg}, \ref{eq:clgk} \& \ref{eq:clgt} and We convert them to equivalent band power before comparing with data. It has the following free parameters: $\theta=\{
A_{\kappa}, A_{\rm ISW}, a^i, x_0^i\}$. $a^i$ and \smash{$x_0^i$} are nuisance parameters to account for uncertainties for our photo-$z$ calibration. We impose $\sum_ix_0^i=0$, where the indices of the redshift bins are $i=0, 1, 2, 3$, and so there are 7 degrees of freedom for the nuisance parameters. $A_{\kappa}$ and $A_{\rm ISW}$ are the key parameters of interest, which characterise the amplitudes of the lensing and ISW signals relative to the fiducial model, as discussed above. All other cosmological parameters are fixed to the \planck\ cosmology. Galaxy bias is a further nuisance parameter, but this will be constrained from data.

\subsection{Galaxy auto- and cross-correlations}\label{sec:results-clgg}

We now present the auto- and cross-correlations from the different tomographic bins. We will use the results to constrain galaxy bias and also to determine the empirical form of the photo-$z$ error distribution.
The galaxy auto-power requires shot noise to be subtracted. Given $N_g$ galaxies in a redshift slice, the shot noise spectrum is given by $C_{\ell}^{\rm shot}=4\pi f_{\rm sky}/N_g$. There is no correction to be made to the cross-power between the different bins. However, we also consider the cross-correlation between our data and that of Z20 and the computation of shot noise is more complicated in that case, since it depends on the numbers of galaxies that are in common to the two catalogues (which is non-zero even for cross-correlation).

Data points with error bars in Fig.~\ref{fig:clgg} show the 10 measured galaxy auto- and cross-correlations for our data. The off-diagonals show the cross-correlation coefficients, defined as
\begin{equation}
    r_{ab}=\frac{C^{ab}_{\ell}}{\sqrt{C^{a}_{\ell}C^{b}_{\ell}}},
\end{equation}
where $a$, $b$ refers to different redshift slices. 
These are independent of galaxy bias.

In this procedure of finding photometric redshift errors, we use only the large-scale modes with $\ell_{\rm max}=500$ as discussed above. The cross-correlation coefficients are flat over a large range of $\ell$, and is only dependent on the redshift distribution.
Specifically, using constraints from the 10 auto- and cross-correlations of galaxy redshift bins, we compute $\chi^2$'s in the 7D nuisance parameter space [$a^i$ $x_0^i$] for $p(z)$. The fitting also excludes $\ell<10$ modes. We use a 2-bias model, detailed in the next section, to find the best-fit $p(z)$. 

Finally, we note that the use of cross-correlations in calibrating $p(z)$ is potentially problematic because of lensing. Even with perfect redshift selection, some cross-correlation is expected between different tomographic slices because of magnification bias: lensing by the nearer slice will imprint an image of its density fluctuations on the more distant slice. Indeed, \cite{2020JCAP...05..047K} argue strongly that magnification bias should be allowed for in CMB lensing tomography. However, we can see that such effects are unimportant here, as they should be largest for widely separated bins, and where the bin has the largest count slope. This should affect above all bin 3, with the highest mean redshift and the highest count slope (the slopes in slices 0--3 are respectively $s\equiv d\log_{10}N/dm=0.19$, 0.29, 0.41, 0.57). But we see from Fig.~\ref{fig:clgg} that bin 3 has no significant correlation with bins 0 and 1. The reason for our different conclusion regarding magnification bias is that our photo-$z$s are calibrated using the colours of spectroscopic objects, whereas \cite{2020JCAP...05..047K} calibrated their photo-$z$s using the cross-clustering with a spectroscopic sample. Magnification bias can affect that cross-correlation and hence the inferred $p(z)$, but it has no effect on the numbers of objects at a given colour.

\begin{figure}
	\centering
	\includegraphics[width=0.47\textwidth]{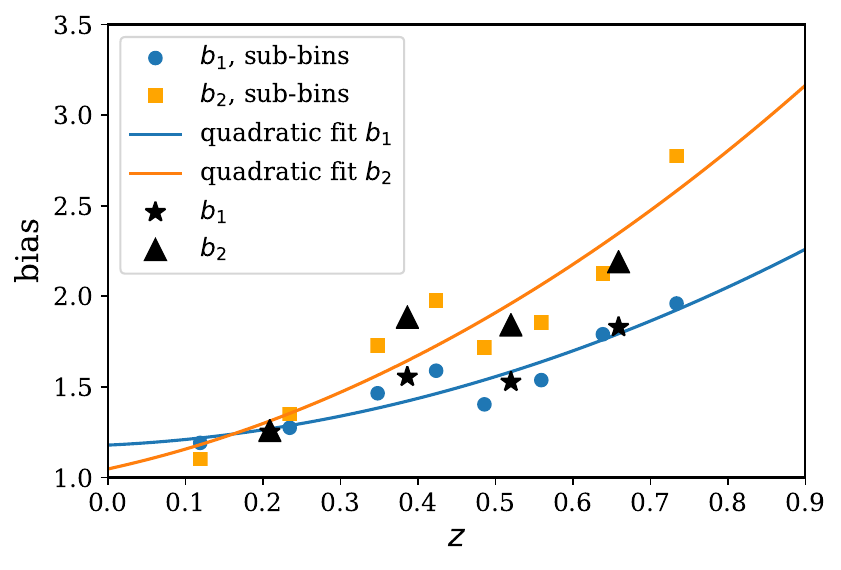}
	\caption{Linear and non-linear bias parameters, $b_1$ and $b_2$ (Eq. \ref{eq:nlbias}), as a function of mean redshift. The circles show minimum-$\chi^2$ bias measured in 8 sub-bins, the stars and triangles show that measured in 4 bins, and the solid lines show quadratic fits to the circles.}
	\label{fig:bias_evolution}
\end{figure}

\begin{figure*}
	\centering
	\includegraphics[width=\textwidth]{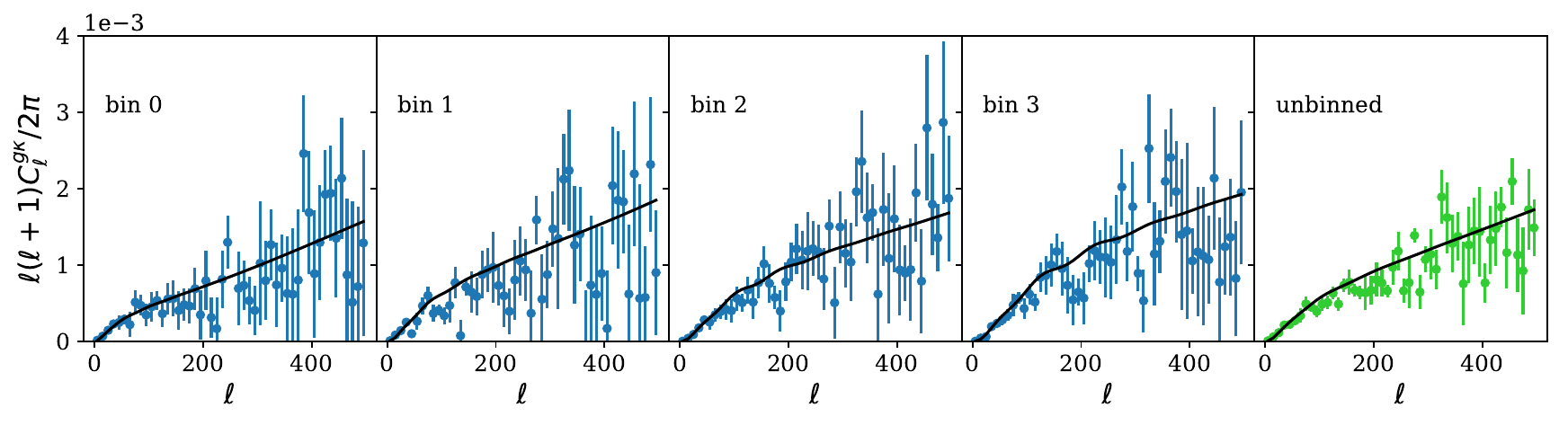}
	\caption{The galaxy-lensing cross-correlation $C_{\ell}^{g\kappa}$ for each redshift slice and the unbinned case. The solid lines are theory with the best-fit $p(z)$ and the same galaxy biases as in Fig.~\ref{fig:clgg}.}
	\label{fig:clgk}
\end{figure*}

\begin{table}
	\centering
	\caption{The effective redshift and the perturbation to the quadratic fits of the bias evolution.}
	\label{tab:bias}
\begin{tabular}{ c|c|c|c|c|c }
 \hline
 Bin & 0 & 1 & 2 & 3 & unbinned \\
 \hline
 $z^{\rm eff}$ & 0.21 & 0.39 & 0.52 & 0.66 & 0.42\\
 $\delta b_1$ & \llap{$-$}0.010 & 0.098 & \llap{$-$}0.033 & 0.029 & \llap{$-$}0.005\\
 $\delta b_2$ & \llap{$-$}0.022 & 0.159 & \llap{$-$}0.056 & \llap{$-$}0.056 & 0.027\\
 \hline
\end{tabular}
\end{table}

\subsubsection{Non-linear bias and bias evolution}

The galaxy auto-power data beyond $\ell\simeq250$ cannot be fit well by a constant bias. Specifically, the ratio between $C_{\ell}^{\rm data}$ and $C_{\ell}^{\rm DM}$ are roughly constant at small and large $\ell$, with a transition at intermediate scales corresponding to roughly the transition between linear and non-linear scales. We allow for this by introducing two bias parameters for the linear and non-linear regimes separately:
\begin{equation}
    C^{gg}_{\ell}=b^2_1 C^{\rm lin}_{\ell}+b^2_2 \Delta C^{\rm nl}_{\ell},
    \label{eq:nlbias}
\end{equation}
where $C^{\rm lin}_{\ell}$ and the nonlinear correction $\Delta C^{\rm nl}_{\ell}$ are computed using the linear and additional non-linear components of the CAMB power spectrum. This simple model gives an excellent fit up to $\ell=1000$. The best-fit linear and non-linear biases using the best-fit $p(z)$ are shown in Table~\ref{tab:Ak_AISW}. We note that $b_2$ is systematically larger than $b_1$, obeying the approximate relation $b_2-1 \simeq 1.9(b_1-1)$.

In the marginalized case, to speed up the computation, we approximate the best-fit biases by taking the ratio of the data with the linear and non-linear theory at different scales using
\begin{equation}
    b_{1,2}^2=\sum_{\ell}w(\ell)\frac{C_{\ell}^{\rm data}}{C_{\ell}^{\rm th}}, \quad w(\ell)=\frac{1/\sigma^2_{\ell}}{\sum_{\ell}(1/\sigma^2_{\ell})}.
    \label{eq:biasratio}
\end{equation}
The transition scales are different for each redshift slice. For bias fitting, a good approximation is the scale at which the fraction of the nonlinear power becomes comparable to the measurement error. This ranges between $\ell\sim100-200$ from low to high redshift slices.
The drawback of this approximation is that the intermediate scales are hard to control, but it gives biases close to the lowest $\chi^2$ value. In this case, the best-fit $p(z)$ gives $\chi^2=471$ with ${\rm DOF}=483$. The model parameters are shown in Table~\ref{tab:redshift_error}.
The best-fit spectra are shown as black solid lines in Fig.~\ref{fig:clgg}, with the galaxy biases and break-down of $\chi^2$ printed for each case. The measured galaxy biases and their errors for each redshift slice are shown in Table~\ref{tab:Ak_AISW}. We have checked that with $\ell_{\rm max}=500$, the best-fit $p(z)$ model and the marginalized case give almost identical amplitude constraints on the cross-correlation of CMB lensing and ISW effects. Therefore, in the following analysis, we will carry out the modelling using the best-fit $p(z)$.

The linear and non-linear biases evolve with redshift, with $b_1$ increasing from $1.2$ to $2.0$ over redshift $0.2$ to $0.7$, although the trend is not quite monotonic (see Fig.~\ref{fig:bias_evolution}). This is consistent with the expectation for luminosity-limited galaxy samples in which high-$z$ galaxies are intrinsically brighter, thus those galaxies tend to occupy more massive dark matter haloes. In general, such evolution can be locally treated as a constant if the redshift bin is thin, or if the distribution is symmetric. However, for bin 3, which has a tail towards higher redshifts, and for an analysis of the unbinned sample, such an approximation breaks down, and the full bias evolution needs to be included in the kernel. To determine the bias evolution more precisely, we sub-divided each bin into two bins. We approximate the redshift distribution of each sub-bin by convolution of the raw $p(z)$ with the best-fit photo-$z$ error of that bin. Then for each sub-bin we fit linear and non-linear biases as above. These measurements are consistent with the 4-bin case. The biases as a function of the mean redshift in that bin can be fitted by a quadratic function (see Fig.~\ref{fig:bias_evolution}). We only use the increasing part of the quadratic, and extrapolate the decreasing part beyond the function's minimal point by a constant.
To match the auto-correlation amplitude, for each bin, we introduce a small correction \smash{$b_i(z)=(1+\delta b_i)b_i^0(z)$}, where $i=1,2$, $b_i^0(z)$ is the fitted quadratic curve, and $\delta b_i\ll 1$. We find $\delta b_i$ by iteration, shown in Table~\ref{tab:bias}.

This model agrees with our measurements very well in general, as seen in Fig.~\ref{fig:clgg}, with reasonable $\chi^2$/DOF overall and for most individual spectra. The auto-power for bin 1 has $\chi^2$ on the high side, but we were unable to identify any systematics that could account for this (e.g. looking for discrepant sky sub-areas in the data). In any case, the look-elsewhere effect is clearly relevant here, with 10 spectra to consider. It is worth noting that $\chi^2$ is nominal for bin 3, even though this has the largest volume and the lowest errors. Indeed, the precision of this bin and bin 2 is sufficient to show a clear signal from  Baryon Acoustic Oscillations (BAO).

Overall, then, these cross-correlation results reassure us that the clustering of the galaxy samples and the calibration of the underlying $p(z)$ distributions is robust, and that the samples are ready for the cross-correlation analysis with the CMB.

\begin{figure}
	\centering
	\includegraphics[width=0.47\textwidth]{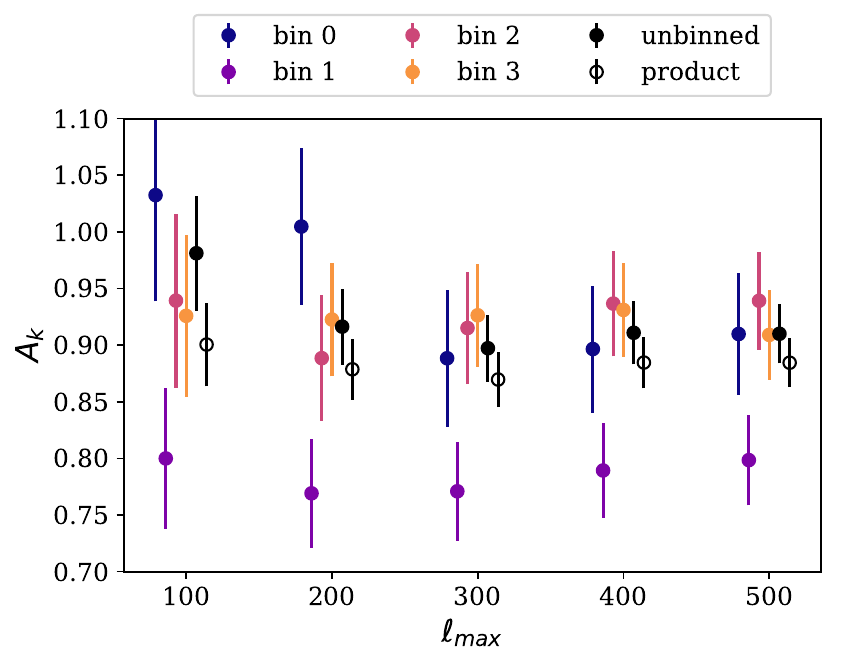}
	\caption{The mean and 1-$\sigma$ of $A_{\kappa}$ likelihoods. Individual bins are shown in blue (bin 0), purple (bin 1), pink (bin 2), and orange (bin 3) points, while the product of the four bins is shown in black open circles. The solid black points show the unbinned case, using the set of best-fit $p(z)$.}
	\label{fig:A_k}
\end{figure}

\begin{figure*}
	\centering
	\includegraphics[width=\textwidth]{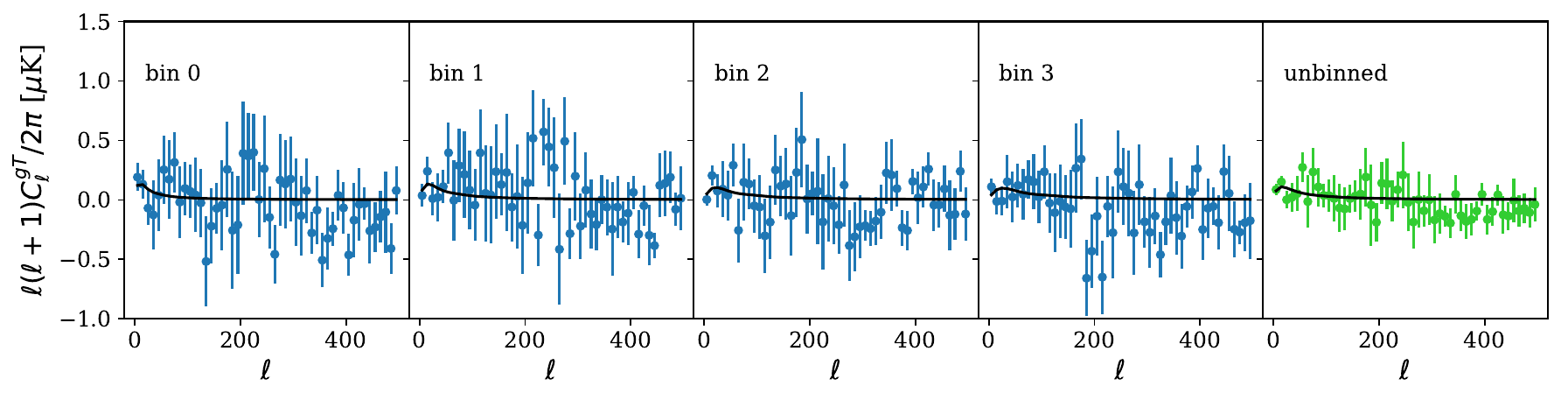}
	\caption{The galaxy-temperature cross-correlation $C_{\ell}^{gT}$ for each redshift slice and unbinned case. The solid lines are the fiducial $\Lambda$CDM predictions with the best-fit $p(z)$ and the same galaxy biases as in Fig.~\ref{fig:clgg}.}
	\label{fig:clgt}
\end{figure*}

\subsection{Galaxy-lensing cross-correlations}
\label{sec:glx}

In computing the galaxy-lensing cross-power signal, we encountered unexpected practical issues. The \planck\ CMB lensing data are made available as spherical harmonic coefficients, from which the required $\kappa$ map can be obtained by using the {\tt alm2map} routine within the {\tt healpy} package. The maximum wavenumber is 2048 in the 2015 release and 4096 in the 2018 release. The 2015 map is already dominated by small-scale noise, but the noise spectrum in the 2018 map displays a nearly divergent spike at high $\ell$: \smash{$C_\ell^\kappa$} increases from about $10^{-4}$ at $\ell=3650$ to over unity at $\ell=4096$. This creates numerical problems in reconstructing the map, so that e.g. making a map at $N_{\rm side}=512$ directly yields a different answer to creating a map at 2048 and downgrading to 512. The spike at $\ell=2048$ can be tamed by filtering the map, but a sufficiently large FWHM is required that modes at $\ell<100$ would be affected. In practice, therefore, we chose to truncate the data at $\ell=2048$, consistent with the 2015 data. With the adoption of a standard resolution of $N_{\rm side}=1024$ for our analysis, the results were robust (and only slightly different from $N_{\rm side}=512$).


A further issue concerned coordinate systems: the CMB maps are supplied in galactic coordinates, whereas we constructed our galaxy maps in equatorial coordinates. Facilities exist within {\tt healpy} for performing the rotation
in $a_{\ell m}$, but we found that the rotation generated artefacts in the lensing auto-power $C^{\kappa}_{\ell}$, which we attribute to the extremely noisy nature of the lensing map, dominated by fluctuations on the inter-pixel scale. After tests at a range of resolutions, we are confident that this issue does not affect the regime of our measurements, out to $\ell=500$. 

Fig.~\ref{fig:clgk} shows the measured galaxy-$\kappa$ cross-power, with the solid black lines showing the theory using the best-fit $p(z)$ and biases obtained from the galaxy auto- and cross-correlations. The black lines are not fits to the data points. To quantify the consistency between data and theory, we include a scaling factor for the lensing amplitude, $A_{\kappa}$, such that $C_{\ell}^{\rm th}=A_{\kappa}bC_{\ell}^{\rm DM}$. 
The constraints on $A_{\kappa}$ as a function of maximum $\ell$-mode is shown in Fig.~\ref{fig:A_k}. The coloured points show measurements from individual tomographic slices, the black open circle shows the product of the four likelihoods, and the black solid points show that from the unbinned case. The mean and 1-$\sigma$ deviation for each of these likelihoods are presented in Table~\ref{tab:Ak_AISW}.

\begin{figure}
	\centering
	\includegraphics[width=0.47\textwidth]{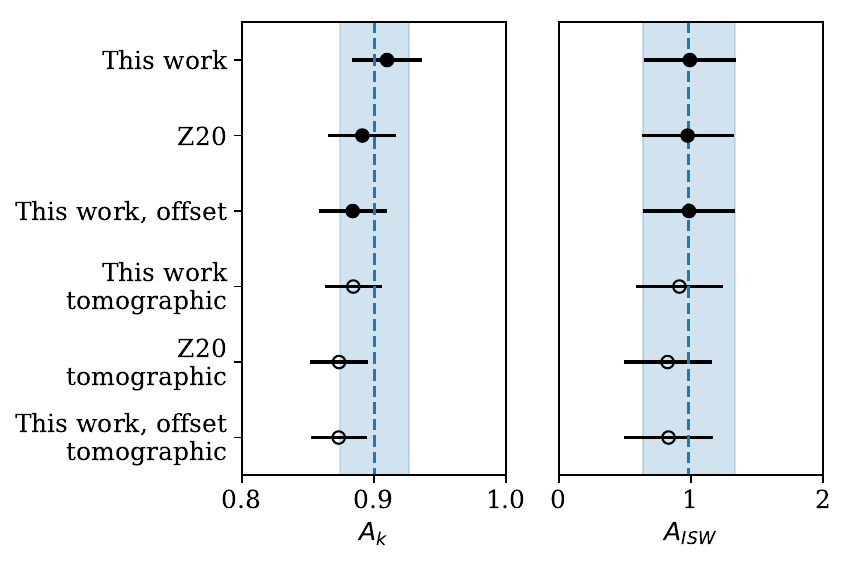}
	\caption{Measurements of $A_{\kappa}$ and $A_{\rm ISW}$ for various data selections at $\ell_{\rm max}=500$ using the appropriate best-fit $p(z)$ for each set. The blue dashed line and band shows our default result, which is the average of the first two data points in each column. These represent a single unbinned analysis, as opposed to the average of the results for the various tomographic shells. The `offset' results refer to the impact of the mean differences between our photo-$z$s and those of Z20 (see Appendix \ref{app:A}).}
	\label{fig:systematic}
\end{figure}

A tendency for the CMB lensing signal to lie below the fiducial model is seen consistently in all tomographic bins. It is also a robust feature, which does not alter with different treatments of the photometric redshifts. We summarise the results of a number of options that we considered in Fig.~\ref{fig:systematic}. We can consider our photometric redshifts or those of Z20; we can further restrict the Z20 sample to objects in the calibratable region of multicolour space; we can clip the photo-$z$ catalogues to remove objects where the estimates are discrepant (we choose a threshold of $|\Delta z|=0.05$); we can adjust one of the photo-$z$ catalogues to remove any offset in $\langle\Delta z\rangle$ as a function of redshift; we can remove objects that are placed in different tomographic bins by the two catalogues. All of these options potentially alter the error distribution and hence the true $p(z)$ of the selection. The nuisance parameters governing these distributions were therefore re-optimised using the galaxy cross-correlations in each case. The impact of some of these
different choices is shown in Fig.~\ref{fig:systematic}.

Our conclusion is that all of these options consistently yield $A_\kappa$ close to 0.9, and that the deviation from the fiducial \planck\ prediction is real. In order to report an overall amplitude for $A_\kappa$, we need to combine the different redshift slices, which we do in the simple approximation that the slices are independent. Because this is not exact, we also consider an unbinned analysis in which all objects at $z<0.8$ are combined; this gives closely consistent results to the outcome of averaging the four slices. We adopt the mean of the unbinned measurements using the two sets of photo-$z$s as our final result: 
\begin{equation}
A_\kappa = 0.901 \pm 0.026.
\end{equation}
This significant discrepancy with the fiducial model is one of the principal results of this paper, and we consider its implications below in Section \ref{sec:interp}. A particularly interesting point is that the overall amplitude of CMB lensing, dominated by LSS at $z\simeq2$, is nevertheless consistent with the fiducial model. What we will show is that these two observations in combination require a matter density lower than the fiducial value.

\subsection{Galaxy-temperature cross-correlations}

\begin{figure}
	\centering
	\includegraphics[width=0.45\textwidth]{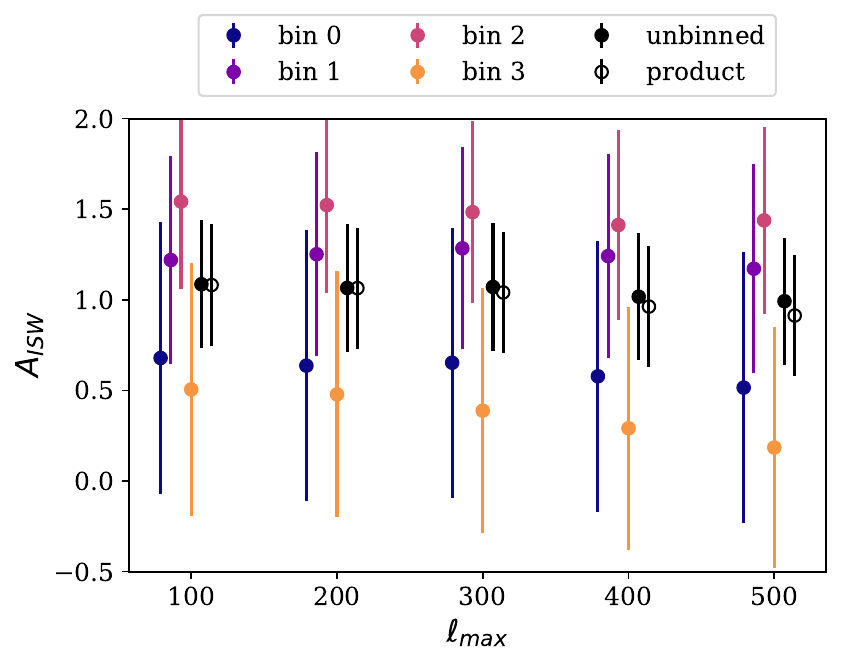}
	\caption{The mean and 1-$\sigma$ of $A_{\rm ISW}$ likelihoods using the set of best-fit $p(z)$. The colour coding and line styles are the same as in Fig.~\ref{fig:A_k}.}
	\label{fig:A_isw}
\vspace{-0.3cm}
\end{figure}

Fig.~\ref{fig:clgt} shows the measurements of galaxy-temperature cross-correlations. The signal is dominated by noise at $\ell>50$. The black solid line shows the theory prediction using the best-fit $p(z)$ and bias from galaxy auto-correlations. 
As with the lensing case, we introduce an ISW amplitude $A_{\rm ISW}$ in order to compare theory and data, such that $C_{\ell}^{\rm th}=A_{\rm ISW}bC_{\ell}^{\rm DM}$. The likelihood for $A_{\rm ISW}$ is then computed for each set of $p(z)$, then marginalized over. The marginalized likelihood for $A_{\rm ISW}$ is almost identical with that of the best-fit model, as shown in Table~\ref{tab:Ak_AISW}. Fig.~\ref{fig:A_isw} shows the likelihoods of $A_{\rm ISW}$ for each redshift slice (coloured) and combined (black) in the marginalized (solid line), mean parameter (circles), and best-fit (dotted line) model cases. The mean and width of individual curve are presented in Table~\ref{tab:Ak_AISW}. The combined likelihood shows a clear detection of the ISW signal, with $A_{\rm ISW}=0.984\pm0.349$, excluding zero at $2.8\sigma$.

In contrast to the CMB lensing signal, the temperature cross-correlation is thus in good agreement with the fiducial $\Lambda$CDM prediction of the ISW effect, although the intrinsically greater cosmic variance on the ISW signal means that we cannot exclude discrepancies at the same level as seen in the lensing signal. The overall modest $S/N$ also prevents strong statements about the signal as a function of redshift, although $A_{\rm ISW}$ is positive and consistent with unity in all bins. The lowest signal is seen in our highest-redshift bin, $A_{\rm ISW}=0.18\pm0.67$, which is interesting in the light of the report by \cite{vst_isw2020} of a null signal at $z=0.68$ using a combined VST+SDSS sample of LRGs: $A_{\rm ISW}=-0.89\pm 0.82$. Our signal is certainly closer to the fiducial $A_{\rm ISW}=1$ than to this result, but the lack of a clear ISW signal in this bin remains.

\begin{table*}
 	\caption{The linear and non-linear bias and constraints on $A_{\kappa}$ and $A_{\rm ISW}$ for various cases at $\ell_{\rm max}=500$. The first row shows the case where all $p(z)$ parameters are marginalized over. The second row shows the case for best-fit $p(z)$ parameters. The third and fourth rows show the cases using the photo-$z$ from \protect\cite{Zhou2020} (Z20) and that with the applied offset. The last row shows the case of using the AvERA model described in \protect\cite{Beck2018}.}
	\centering
	\begin{tabular}{|c| c| c| c| c| c| c|} 
	\toprule
	 {\bf Parameters} & bin0 & bin1  & bin2  & bin3 & combined & Un-binned  \\ [0.8ex] 
	 Redshift & $0 < z \leq 0.3$ & $0.3 < z \leq 0.45$ & $0.45 < z \leq 0.6$ & $0.6 < z \leq 0.8$ & - & $0 < z \leq 0.8$ \\ \hline
	 \multicolumn{7}{|l|}{\bf Marginalized over $p(z)$} \\ 
	$b_1$ & $1.25\pm 0.01$ & $1.53\pm0.02$ & $1.54\pm0.01$ & $1.86\pm0.02$ & - & - \\ [0.9ex]
	$b_2$ & $1.27\pm 0.01$ & $1.85\pm0.03$ & $1.82\pm0.01$ & $2.23\pm0.02$ & - & - \\ [0.9ex]
	$A_k$ & $0.91\pm0.05$ & $0.82\pm0.04$ & $0.94\pm0.04$ & $0.90\pm0.04$ & $0.89\pm0.02$ & - \\ [0.9ex]
	$A_{\rm ISW}$ & $0.52\pm0.78$ & $1.20\pm0.63$ & $1.48\pm0.61$ & $0.18\pm0.67$ & $0.91\pm0.33$ & - \\ [0.9ex] \hline
	 \multicolumn{7}{|l|}{\bf Best-fit $p(z)$} \\ 
	$b_1$ & 1.25 & 1.56 & 1.53 & 1.83 & - & 1.43 \\ [0.9ex]
	$b_2$ & 1.26 & 1.88 & 1.84 & 2.19 & - & 1.59 \\ [0.9ex]
	$A_k$ & $0.91\pm0.05$ & $0.80\pm0.04$ & $0.94\pm0.04$ & $0.91\pm0.04$ & $0.88\pm0.02$ & $0.91\pm0.03$ \\ [0.9ex]
	$A_{\rm ISW}$ & $0.52\pm0.75$ & $1.17\pm0.58$ & $1.44\pm0.52$ & $0.18\pm0.67$ & $0.91\pm0.33$ & $0.99\pm0.35$ \\ [0.9ex] \hline
	  \multicolumn{7}{|l|}{\bf Zhou et. al.} \\ 
	$b_1$ & 1.25 & 1.54 & 1.55 & 1.90 & - & 1.44 \\ [0.9ex]
	$b_2$ & 1.26 & 1.87 & 1.90 & 2.21 & - & 1.62 \\ [0.9ex]
	$A_k$ & $0.91\pm0.06$ & $0.81\pm0.04$ & $0.93\pm0.04$ & $0.87\pm0.04$ & $0.87\pm0.02$ & $0.89\pm0.03$ \\ [0.9ex]
	$A_{\rm ISW}$ & $0.50\pm0.79$ & $1.03\pm0.59$ & $1.37\pm0.55$ & $0.20\pm0.63$ & $0.82\pm0.33$ & $0.98\pm0.35$ \\ [0.9ex] \hline
	   \multicolumn{7}{|l|}{\bf Offset} \\ 
	$b_1$ & 1.28 & 1.52 & 1.54 & 1.89 & - & 1.45 \\ [0.9ex]
	$b_2$ & 1.30 & 1.86 & 1.87 & 2.20 & - & 1.64 \\ [0.9ex]
	$A_k$ & $0.89\pm0.05$ & $0.81\pm0.04$ & $0.93\pm0.04$ & $0.89\pm0.04$ & $0.87\pm0.02$ & $0.88\pm0.03$ \\ [0.9ex]
	$A_{\rm ISW}$ & $0.45\pm0.81$ & $1.05\pm0.58$ & $1.32\pm0.56$ & $0.25\pm0.46$ & $0.83\pm0.33$ & $0.99\pm0.35$ \\ [0.9ex] \hline
	\multicolumn{7}{|l|}{\bf AvERA model} \\ 
	$b_1$ & 1.16 & 1.34 & 1.25 & 1.46 & - & 1.23 \\ [0.9ex]
	$b_2$ & 1.11 & 1.50 & 1.45 & 1.75 & - & 1.33 \\ [0.9ex]
	$A_k$ & $0.97\pm0.06$ & $0.80\pm0.04$ & $0.91\pm0.04$ & $0.85\pm0.04$ & $0.87\pm0.02$ & $0.91\pm0.03$ \\ [0.9ex]
	$A_{\rm ISW}$ & $0.24\pm0.35$ & $0.48\pm0.25$ & $0.55\pm0.23$ & $0.07\pm0.24$ & $0.35\pm0.13$ & $0.39\pm0.14$ \\ [0.9ex]
	 
\bottomrule 
	\end{tabular}
	\label{tab:Ak_AISW}
\end{table*}

\section{Implications for the cosmological model}\label{sec:interp}

\subsection{Implications of low $A_\kappa$}

We first consider the simplest interpretation of our low $A_\kappa$ amplitude for the galaxy-CMB lensing cross correlation in terms of parameters within the $\Lambda$CDM model. The lensing signal at low $z$ has a direct linear dependence on the matter density fluctuation, which is proportional to the mean density times the relative fluctuation -- i.e. to $\Omega_m\sigma_8$. The cross-correlation is also proportional to galaxy bias, but we have shown in Section \ref{sec:glx} how that degree of freedom can be determined separately by including the galaxy auto-correlation data. At non-zero redshifts, the dependence on $\Omega_m$ becomes nonlinear as this parameter influences distances and evolution of density fluctuations. For our range of redshifts, the empirical density dependence of the amplitude is as $\Omega_m^{0.78}$, so that our result for $A_\kappa$ produces the following constraint:
\begin{equation}
    \sigma_8\Omega_m^{0.78} = 0.297 \pm 0.009.
\end{equation}
It is interesting to note that total CMB lensing itself produces a constraint of a similar form, but with a different density dependence:
\begin{equation}
    \sigma_8\Omega_m^{0.25} = 0.589 \pm 0.020
\end{equation}
\citep{PlanckLens2018}.
A straightforward combination of these two results yields
\begin{equation}
    \Omega_m = 0.275 \pm 0.024;\quad
    \sigma_8 = 0.814 \pm 0.042;
\end{equation}
the same normalization as \planck, but a somewhat lower density.

\begin{figure}
	\centering
	\includegraphics[width=0.47\textwidth]{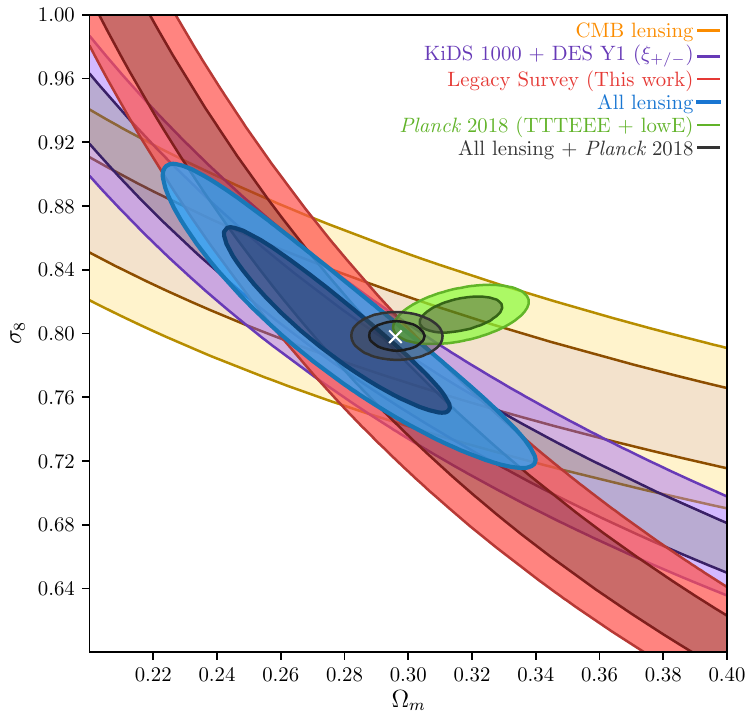}
	\caption{Comparison of constraints on the $\Omega_m-\sigma_8$ plane; the contours contain 68\% and 95\% of the total probability. Note the consistent intersection of the three lensing-based results. The KiDS-1000 + DES Y1 constraint are for cosmic shear only. We use a Gaussian likelihood for DES Y1 and a skewed Gaussian using model 2 of \protect\cite{2003physics...6138B} for KiDS-1000 to account for their asymmetric errors.
	}
	\label{fig:oms8}
\end{figure}

It is interesting to compare these results with analogous constraints from weak galaxy lensing. Here the dependence on density is intermediate in strength. The constraints from the cosmic shear measurement of KiDS-1000 \citep{2020arXiv200715633A} and DES Y1 \citep{2018PhRvD..98d3528T} are as follows:

\begin{align}
    \sigma_8\Omega_m^{0.5} & = 0.416^{+0.013}_{-0.011}
    \qquad {\rm KiDS-1000 } \\
    \sigma_8\Omega_m^{0.5} & = 0.428 \pm 0.015 \qquad {\rm DES-Y1}
\end{align}
which is in close consistency with what would be deduced from the CMB lensing results: $\sigma_8\Omega_m^{0.5}=0.427$, as opposed to the fiducial 0.455. In combination, these three lensing results then give a clear preference for a model with a rather lower density than the \planck\ fiducial model, as illustrated in Fig.~\ref{fig:oms8}:
\begin{equation}
    \Omega_m = 0.274 \pm 0.024;\quad
    \sigma_8 = 0.804 \pm 0.040.
\end{equation}
It can be noted that the KiDS-1000 papers preferred to interpret their results in terms of a reduced $\sigma_8$, but a shift purely in normalization is disfavoured by the total CMB lensing amplitude, quite apart from our current results.

The conflict of this result with \planck\ is marked: $\Delta\chi^2=12$ on 2 degrees of freedom, which represents a $p$ value of $2\times 10^{-3}$.  In these circumstances, we should be cautious in accepting the formal combination of the above lensing result with \planck, which is
\begin{equation}
    \Omega_m = 0.296 \pm 0.006;\quad
    \sigma_8 = 0.798 \pm 0.006.
\end{equation}
In fact, this unimaginative compromise model is arguably not ridiculous: it lies within the 95\% confidence contours of both our combined lensing result and \planck. The value of $\Omega_m$ alone would represent a $2.7\sigma$ deviation from \planck, but consistency in $\sigma_8$ is better and there is no prior reason to be more concerned about a deviation in one or other of these parameters. Nevertheless, agreement this weak is asking a lot of bad luck: we may be fairly sure that systematics are present, and the question is whether they lie in the data or in the theoretical framework. From the point of view of the actual values of $\Omega_m$ and $\sigma_8$, `new physics' counts as just a further systematic on the same footing as data errors \citep{BernalPeacock2018}, but these alternatives are hardly equal in their implications.

The most conservative assumption is that there are indeed imperfections in the data. If these were to lie on the lensing side, we would point the finger of suspicion at photometric redshifts, which are a dominant source of potential bias. We have discussed the reliability of the photo-$z$s used in this paper at some length, and a huge effort has been invested in this topic by galaxy weak lensing groups -- using rather different data and methods to the approach taken here. On the CMB side, the exemplary thoroughness of the \planck\ analysis makes one reluctant to suggest any imperfection, but there are issues. Apart from the continuing puzzle of the well-known large-scale anomalies, there is the fact that the \planck\ TTTEEE data prefer an amplitude of CMB lensing that is {\it higher\/} than fiducial:
$A_{\rm lens}=1.180\pm  0.065$, which represents some form of internal inconsistency. These issues were investigated in detail by \cite{GPE_Gratton2019}, who argued that that the anomalies represented no more than slightly unusual statistical fluctuations in the \planck\ dataset and that there was no evidence of systematics at an important level. Where large-scale properties of the CMB are concerned, cosmic variance dominates and the many independent \planck\ maps can indeed give confidence that systematics are negligible. But in the noise-dominated regime, where the best results require the combination of all data subsets, residual systematics at the few-$\sigma$ level are not so easy to rule out. The \planck\ constraint on $\Omega_m$ does depend significantly on the high$-\ell$ data, and so could be considered potentially less robust. We therefore 
think it is plausible that the compromise solution with $\Omega_m\simeq0.296$ may be close to the truth. If we look at CMB constraints independent of \planck, ACT+WMAP yields $\Omega_m=0.313\pm 0.016$, which is easily consistent with 0.296; this work also has $A_{\rm lens}$ very close to unity \citep{ACTpars2020}.

\subsubsection{Implications for the Hubble parameter}

A slightly reduced matter density would also have the advantage of reducing the other tension that is currently the subject of much discussion: the Hubble parameter. The most robust inference concerning $H_0$ from the CMB comes from the main acoustic scale, which can be taken empirically as measuring the combination $\Omega_m h^3$ with negligible error. If we use this as a basis for rescaling the fiducial model, the compromise $\Omega_m=0.296$ would require $H_0\simeq 69\kmsmpc$. This 2\% increase from the fiducial value is still significantly below the direct determination of $74.03\pm1.42$ \citep{Riess2019}, but again would only require a modest level of systematics for consistency. Furthermore, taking seriously the $\Omega_m\simeq0.274$ from the combined lensing data would imply a completely consistent $H_0\simeq 71$.

Consideration of variations in $h$ prompts us to ask whether the predicted $A_\kappa$ depends on $h$. From Eq.~\ref{eq:clgk}, we can see that there is no explicit $h$ dependence, since $h$ times comoving distance is a function of redshift and $\Omega_m$ only. The scale at which $\sigma_8$ is determined is accessible to the range of $\ell$ under study, so changes in power-spectrum shape arising from changes in $h$ would be expected to have a minor effect. In practice, we find $A_\kappa\propto h^{0.24}$, which is equivalent to a negligible $\Omega_m^{0.08}$ effect when considering variations with $\Omega_mh^3$ fixed.

It is undeniably depressing to be considering the possibility that one or more of the leading current cosmological datasets could be reporting results that contain systematic errors of close to $2\sigma$, but equally we need to beware of too hastily declaring the existence of new physics as soon as we see a minor statistical discrepancy. Because there are in principle two distinct discrepancies, affecting $\Omega_m-\sigma_8$ and $H_0$, a single new addition to the cosmological model that solved both issues would demand to be taken seriously. But both the lensing and $H_0$ discrepancies have existed in the literature for some while, and it is fair to say that no compelling solution has emerged. Nevertheless, it is worth reviewing some selected candidates.

\subsubsection{Massive neutrinos}

It is known that neutrinos make a non-zero contribution to the non-relativistic density, with a summed mass of at least $0.06$\,eV
($\Omega_\nu h^2 > 0.00064$). Owing to free streaming, the neutrino distribution is close to homogeneous on the scales of LSS, and therefore the lensing effect is reduced in two ways: the clumped mass is only the CDM, with a density $(1-f_\nu)\Omega_m$; this lower effective dark matter density slows growth since last scattering, reducing $\sigma_8$ today. At first sight, these effects sound as if they have the potential to close the gap between lensing results and \planck, but this is not so. Firstly note that we do not really need to be concerned with growth suppression for the interpretation of the lensing results themselves, since the lensing signal is directly proportional to the low-redshift normalization. Furthermore, the standard definition of $\sigma_8$ (adopted by \planck\ and CAMB) is that it is the rms fractional fluctuation in the {\it total\/} matter density. The fractional fluctuation in the CDM density is thus $\sigma_8/(1-f_\nu)$, and this raised amplitude compensates for the lower clumped density, so that the lensing signal for a given $\Omega_m$ and $\sigma_8$ should be independent of the neutrino fraction. The only subtlety is that the growth between $z=1-2$ and $z=0$ will be slightly less than in $\Lambda$CDM for the given $\Omega_m$. But this is a tiny effect: $f_g$ is about $\Omega_m(z)^{0.6}$, so the relative $f_g$ is $(1-f_\nu)^{0.6}$, so the mass fluctuations at $z=1-2$ are higher by of order $1+0.6f_\nu$ than in $\Lambda$CDM for a given $z=0$ normalization, which is a negligible correction.

Therefore, all the dependence on neutrino fraction on the $\Omega_m-\sigma_8$ plane comes from \planck. Inspecting their chains, the effect is approximately $\sigma_8 \propto (1-f_\nu)^{2.2}$  and $\Omega_m \propto (1-f_\nu)^{-2.5}$. Although the predicted normalization is reduced, as expected, the best-fit density rises and so the tension between primary CMB and lensing is {\it increased\/} if there is a non-minimal neutrino fraction.

\subsubsection{Modified gravity}

A more effective modification of theory concerns the strength of gravity. To avoid excessive complication, it is common to approach this in a form that includes two linear parameters that modify the scalar potentials $\Psi$ and $\Phi$, which describe fluctuations in the time and spatial parts of the metric. In the standard model, $\Psi=\Phi$ and the potentials satisfy the Poisson equation. The most transparent modification is to scale the forces for non-relativistic particles
(from $\Psi$) and photons (from $\Psi+\Phi$) that result from a given mass fluctuation, $\delta$, so that $\nabla^2\Psi\propto (1+\mu)\delta$ and $\nabla^2(\Psi+\Phi)\propto (1+\Sigma)\delta$ (e.g. \citejap{Simpson2013}). The motivation for modified gravity comes from late-time accelerated expansion, and therefore it is normally assumed that the modifications evolve as
\begin{equation}
    (\mu(z),\Sigma(z)) = (\mu_0,\Sigma_0)\,\Omega_\Lambda(z),
\end{equation}
so that modifications are unimportant at last scattering. Since $\Lambda$CDM seems to describe the expansion history well, it is also assumed that the modifications affect only perturbations. Thus the cosmological parameters inferred from the CMB should be unaffected in this framework, and therefore modified gravity can be used to close any gap between the predicted and observed lensing signal. There is a degeneracy here: for $\Sigma=0$ (normal lensing strength), we can appeal to $\mu<0$ to reduce the growth in fluctuations; alternatively, we can have normal growth with $\mu=0$ and suppress the resulting lensing signal by appealing to $\Sigma<0$. In either of these solutions, it would be understandable that the total CMB lensing signal is consistent with standard gravity, because it arises around $z=2$, where the modifications are only just switching on. To achieve $A_\kappa\simeq0.9$ at $z\simeq0.5$, where $\Omega_\Lambda=0.4$, we need either $\Sigma_0=-0.25$, or $\mu_0=-1.5$. The large value for $\mu_0$ seems surprising at first sight, implying close to total suppression of LSS gravity at the present epoch. This is partly a consequence of the $\mu\propto\Omega_\Lambda(a)$ assumption, and also because $\mu$ suppression of the strength of gravity only alters the growth {\it rate\/}: to achieve significant reduction in $\delta$ at $z\simeq 0.5$ would require substantial alteration to the growth rate at much higher redshifts, which is hard to achieve in this model unless $\mu_0$ is large. Such a model can be ruled out by other evidence, since it would imply a very non-standard growth rate at $z=0.5$, whereas we know from redshift-space distortions that the rate is within about 10\% of fiducial at this redshift \citep{eboss2020}.

In summary, then, an explanation of a low lensing amplitude via modified gravity must involve an alteration of the strength of light deflection by a given mass concentration, rather than reducing the amplitude of mass fluctuations. Such an explanation appears to be consistent and not in conflict with other evidence, but one could hardly call it compelling -- not least because it has no impact on the $H_0$ tension; such a radical conclusion requires more than a single piece of evidence. In due course, we will have more accurate tomographic lensing and redshift-space distortion data where changes in the growth rate and strength of lensing with redshift can be measured, so that a progressive decline in the strength of lensing could be measured. Without such evidence, this hypothesis is at best provisional.

\begin{figure*}
	\centering
	\includegraphics[width=\textwidth]{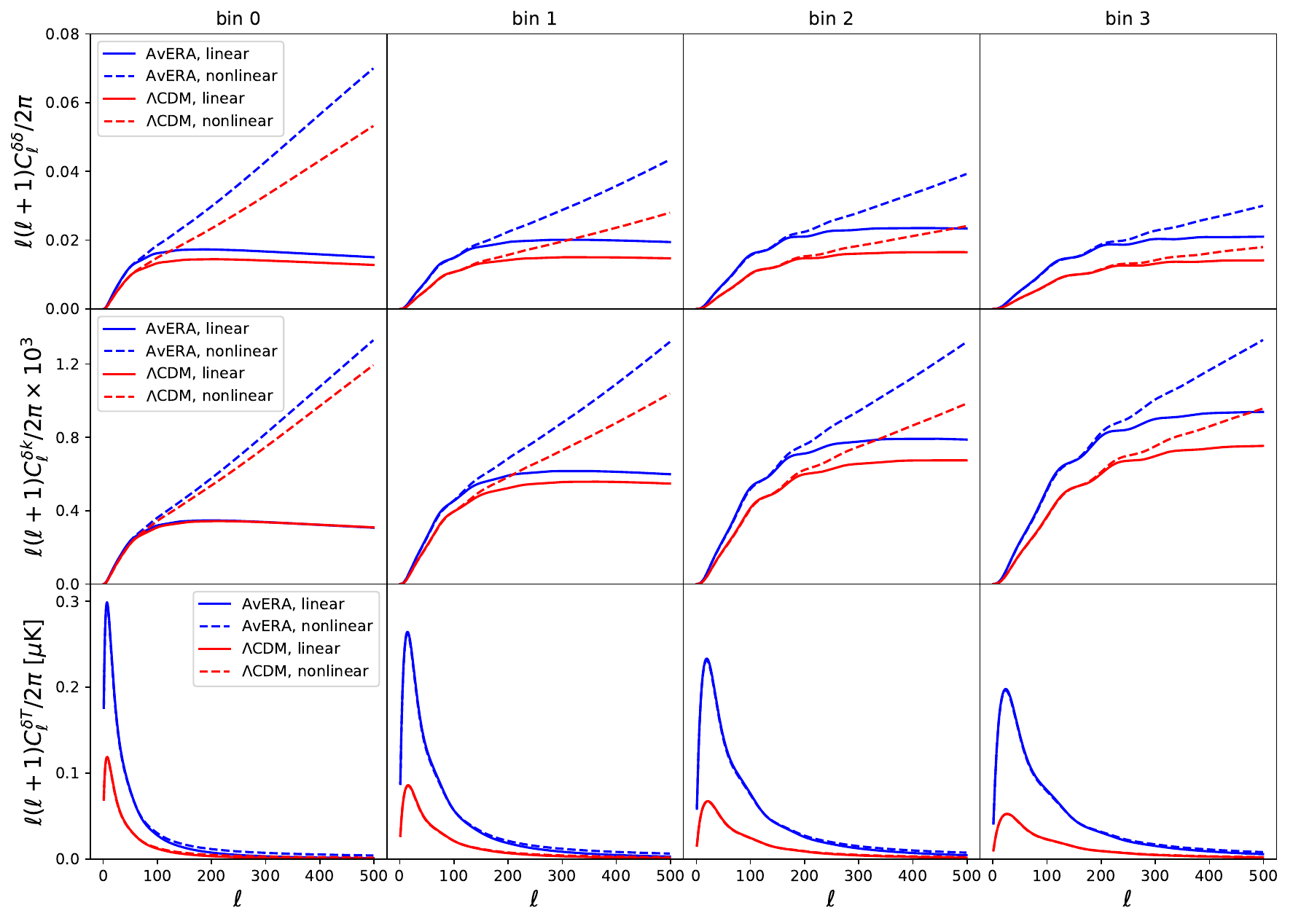}
	\caption{The dark matter auto-correlation (top), the matter-$\kappa$ cross-correlation (middle), and the matter-temperature cross-correlation (bottom) in $\Lambda$CDM (red) and AvERA (blue) model for the four tomographic bins using the best-fit $p(z)$. The solid lines show computation using linear power spectrum, and the dashed lines show that using non-linear power spectrum from {\sc Halofit}.}
	\label{fig:avera_cl}
\end{figure*}

\begin{figure}
	\centering
	\includegraphics[width=0.47\textwidth]{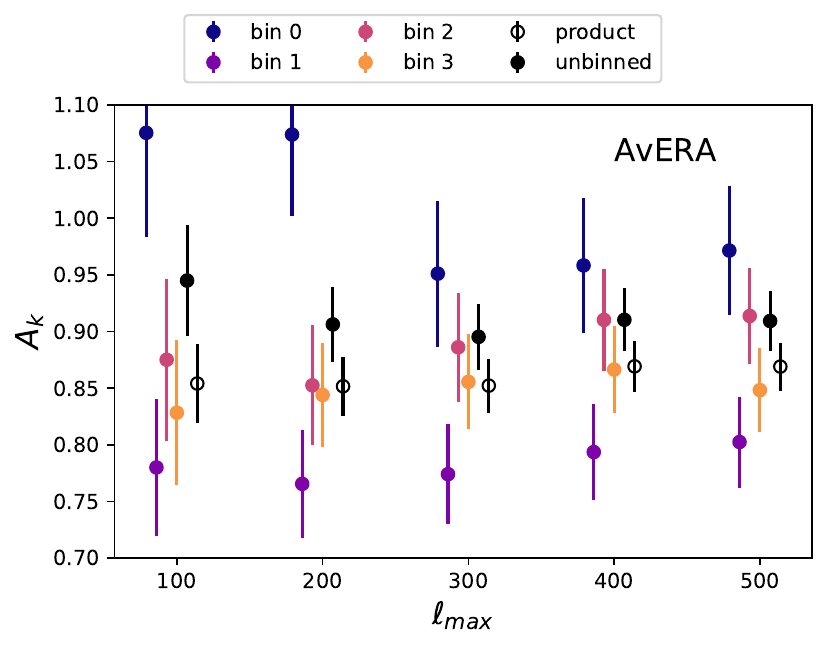}
	\caption{The constraints for $A_{\kappa}$ from the normalized likelihoods in the AvERA model using the best-fit $p(z)$ and fitted galaxy bias.}
	\label{fig:avera_Ak}
\end{figure}

\begin{figure}
	\centering
	\includegraphics[width=0.47\textwidth]{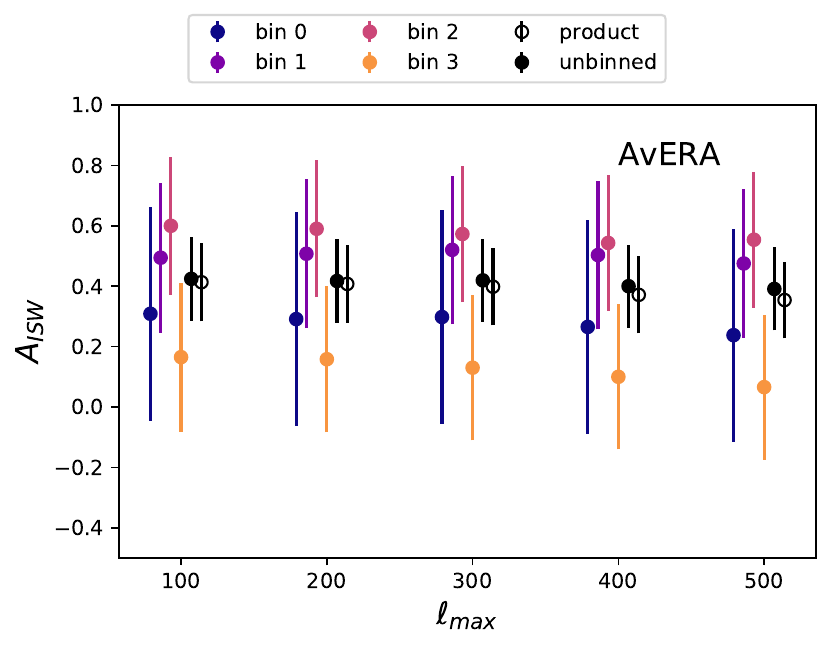}
	\caption{The constraints for $A_{\rm ISW}$ from the normalized likelihoods in the AvERA model using the best-fit $p(z)$ and fitted galaxy bias.}
	\label{fig:avera_At}
\end{figure}

\subsection{ISW and the AVERA model}

An interesting approach that has been proposed with a view to explaining the high claimed ISW signal from superstructures is the AvERA model \citep{avera2017}. This is a radical framework that postulates a critical-density universe without a cosmological constant, but with averaging of an inhomogeneous expansion rate, leading to an apparent acceleration as measured by the mean effective Hubble parameter. The model can be adjusted so that the empirical $H(z)$ relation rather closely matches the standard $\Lambda$CDM case -- which has the advantage that the conversion between distance and redshift remains as in the standard model, so that inferences from the CMB regarding density parameters and the shape of the matter power spectrum remain valid.

On the other hand, the {\it amplitude\/} of the spectrum is modified in this model, and the density growth rate $f_g\equiv d\ln\delta/d\ln a$ is rather different from $\Lambda$CDM. There is a spike above $f_g=1$ around $z\simeq 2$ and in general the rate is higher than the standard model; thus, the required value of $\sigma_8$ at $z=0$ has to be increased in order to be consistent with the amplitude of primordial fluctuations inferred from the CMB. A convenient fitting formula for the growth rate is
\begin{equation}
    f_g(a) = \exp(-2.308a^2) + 0.549 [1+11.569(\ln a + 1.222)^{2}]^{-1}.
    \label{eq:avera_fit}
\end{equation}
Integration of this expression implies that $\sigma_8(z)$ for AvERA is above $\Lambda$CDM at high redshift, by as much as a factor 1.2 at $z=1.5$. Conversely, the low-redshift evolution is slower and the amplitude of present-day matter fluctuations is about 5\% lower than $\Lambda$CDM. The two models predict identical amplitudes at $z\simeq0.08$. Thus, for redshifts relevant for our tomographic data, the AvERA model predicts a higher density fluctuation, so that the predicted amplitude of the linear ISW signature is greater. There will also be a greater degree of nonlinear evolution. We treat this by assuming that the nonlinearities can be estimated in the {\tt Halofit} framework by taking the standard $\Lambda$CDM approach and increasing $\sigma_8(z)$ appropriately. This should be sufficient to indicate how important the increased nonlinearity might be (this will be more of a potential issue for lensing, where even weak lensing can be dominated by nonlinear structures on small enough angular scales).

We use \cite{Planck2018Param} Cosmological parameters, and set the power spectrum of AvERA to be identical to $\Lambda$CDM at $z=8.55$ consistent with \cite{Beck2018}. We use the fitting formula in Eq.~\ref{eq:avera_fit}, and interpolate the AvERA $H(z)$ and $R(z)$ as given by \cite{Beck2018}. Fig.~\ref{fig:avera_cl} shows the matter auto-correlation, matter-$\kappa$, and matter-$T$ cross-correlations in AvERA and $\Lambda$CDM with both linear and non-linear power spectra, using the best-fit $p(z)$. As expected, the AvERA prediction has a higher amplitude than $\Lambda$CDM. The corresponding galaxy biases are significantly smaller in the AvERA case as shown in Table~\ref{tab:Ak_AISW}, but this effect is absorbed in the lensing cross-correlation, resulting in similar 
constraints on $A_{\kappa}$. The likelihoods for $A_{\kappa}$ and $A_{\rm ISW}$ are obtained in Figs~\ref{fig:avera_Ak}-\ref{fig:avera_At}. In this case, we find $A_{\kappa}=0.87\pm0.02$ for the product, and $A_{\kappa}=0.91\pm0.03$ for the unbinned case. In the ISW case, the AvERA prediction is about three times as large as $\Lambda$CDM. The preferred amplitude is $A_{\rm ISW}=0.35\pm0.13$ from the product of tomographic bins, and $A_{\rm ISW}=0.39\pm0.14$ from the unbinned result. Adopting the unbinned case, this ISW result excludes unity at $4.4\sigma$ and we can be confident that the AvERA model greatly over-predicts the general level of ISW fluctuations.


\section{Summary and discussion}\label{sec:summary}

We have performed a tomographic analysis of the cross-correlations between Legacy Survey galaxies and the \planck\ CMB lensing convergence and temperature maps, covering $17\,739\,{\rm deg}^2$. We obtained our own photometric redshifts for the Legacy Survey based on $g-r$, $r-z$, and $z-W_1$ colours, with a precision of $\sigma_z/(1+z)=0.012-0.015$. The galaxy sample is divided into four wide redshift bins between $z=0$ and $z\simeq0.8$. We model errors in photometric redshift with respect to calibration data sets via a modified Lorentzian function, and constrain the tails of the error distribution by requiring consistent prediction of the galaxy cross-correlation signal between different tomographic bins. This modelling incorporates a novel scheme for dealing with scale-dependent bias (Eq.~\ref{eq:nlbias}), in which the linear and nonlinear parts of the matter power spectrum receive independent boosts to their amplitudes.
The consistency of the galaxy clustering and its cross-correlations argues that the galaxy sample from the Legacy survey is robust, and that the properties of the photometric redshifts are understood. 

We then proceeded to evaluate the cross-correlation between the tomographic galaxy maps and the CMB maps of temperature and lensing convergence. The results are compared with the predictions of the fiducial \planck\ cosmological model, marginalizing over the photo-$z$ error parameters with weights given by the likelihood from fitting galaxy auto- and cross-correlations. 

The amplitude for the ISW signal relative to the fiducial prediction is $A_{\rm ISW}=0.98\pm0.35$, consistent with $\Lambda$CDM, as found by previous works, e.g. \cite{ISW2018}. We also explored the AvERA model \citep{avera2017}, which was developed in order to explain the claimed excess signal in the stacked ISW signal in supervoids. We find that in this model, $A_{\kappa}=0.91\pm0.03$, and $A_{\rm ISW}=0.39\pm0.14$, with significantly smaller galaxy biases compared to the $\Lambda$CDM case. Thus, the AvERA model achieves its aim of predicting an enhanced supervoid signal at the price of raising the overall level of ISW power to the point where it is inconsistent with observation, even given the relatively noisy nature of the ISW signal. If the supervoid signal is found to persist in future studies, AvERA cannot be the explanation.

The amplitude of the CMB lensing signal is found to be significantly lower than the prediction of the fiducial \planck\ model, with a scaling factor $A_{\kappa}=0.901\pm0.026$. We note that this lower amplitude is consistent with the results from an analysis of cross-correlation between CMB lensing and a DESI LRG sample based on the Legacy Survey data \citep{Kitanidis2020b}. Our result can be translated into constraints on the parameter combination $\sigma_8\Omega_m^{0.78}=0.297\pm0.009$. The total CMB lensing signal provides an alternative constraint on this plane, of $\sigma_8\Omega_m^{0.25}=0.589\pm0.020$ \citep{PlanckLens2018}, which also represents an amplitude lower than fiducial, although only by $1\sigma$. In combination, these CMB lensing figures prefer a solution with a relatively low matter density of $\Omega_m\simeq0.274$. These CMB lensing results are also in excellent agreement with the value of $\sigma_8\Omega_m^{0.5}$ deduced from weak galaxy lensing \citep{2018PhRvD..98d3528T,2020arXiv200715633A}.  Within the compass of $\Lambda$CDM, the model that does least violence to lensing and CMB data is
\begin{equation}
    \Omega_m=0.296\pm0.006,\quad \sigma_8=0.798\pm0.006,
\end{equation}
and this is consistent with the 95\% confidence ranges from both datasets. It is therefore worth taking seriously the possibility that the true cosmic density is substantially on the low side of the fiducial \planck\ estimate. Such a reduction would also reduce the $H_0$ tension, raising the best-fitting CMB value to around $69\kmsmpc$ -- although this would still imply the existence of systematics in the direct $H_0$ data (see e.g. \citejap{GPE2020}).

We therefore face a situation where at least two of three currently dominant cosmological probes contain unrecognised systematics at the level of a few standard deviations, or the standard model must be extended. The choice between conservatism or revolution is perhaps not so easy in the current circumstances, but the next generation of experiments should settle the question beyond all doubt.

\section{Data Availability}
All of the observational datasets used in this paper are available through the Legacy Survey website \url{http://legacysurvey.org/dr8/}. The codes used in this analysis along with several processed data products can be accessed at \url{https://gitlab.com/qianjunhang/desi-legacy-survey-cross-correlations}.

\section*{Acknowledgements}

We thank Martin White and Catherine Heymans for their valuable comments. 
We are grateful to the referee, George Efstathiou, for a thorough and thought-provoking report, which helped to clarify several issues in the paper.
QH was supported by the Edinburgh Global Research Scholarship and the Higgs Scholarship from Edinburgh University. SA and JAP were supported by the European Research Council under grant number 670193 (the COSFORM project). YC acknowledges the support of the Royal Society through the award of a University Research Fellowship and an Enhancement Award.

The Legacy Surveys consist of three individual and complementary projects: the Dark Energy Camera Legacy Survey (DECaLS; NOAO Proposal ID \# 2014B-0404; PIs: David Schlegel and Arjun Dey), the Beijing-Arizona Sky Survey (BASS; NOAO Proposal ID \# 2015A-0801; PIs: Zhou Xu and Xiaohui Fan), and the Mayall z-band Legacy Survey (MzLS; NOAO Proposal ID \# 2016A-0453; PI: Arjun Dey). DECaLS, BASS and MzLS together include data obtained, respectively, at the Blanco telescope, Cerro Tololo Inter-American Observatory, National Optical Astronomy Observatory (NOAO); the Bok telescope, Steward Observatory, University of Arizona; and the Mayall telescope, Kitt Peak National Observatory, NOAO. The Legacy Surveys project is honored to be permitted to conduct astronomical research on Iolkam Du'ag (Kitt Peak), a mountain with particular significance to the Tohono O'odham Nation.

NOAO is operated by the Association of Universities for Research in Astronomy (AURA) under a cooperative agreement with the National Science Foundation.

This project used data obtained with the Dark Energy Camera (DECam), which was constructed by the Dark Energy Survey (DES) collaboration. Funding for the DES Projects has been provided by the U.S. Department of Energy, the U.S. National Science Foundation, the Ministry of Science and Education of Spain, the Science and Technology Facilities Council of the United Kingdom, the Higher Education Funding Council for England, the National Center for Supercomputing Applications at the University of Illinois at Urbana-Champaign, the Kavli Institute of Cosmological Physics at the University of Chicago, Center for Cosmology and Astro-Particle Physics at the Ohio State University, the Mitchell Institute for Fundamental Physics and Astronomy at Texas A\&M University, Financiadora de Estudos e Projetos, Fundacao Carlos Chagas Filho de Amparo, Financiadora de Estudos e Projetos, Fundacao Carlos Chagas Filho de Amparo a Pesquisa do Estado do Rio de Janeiro, Conselho Nacional de Desenvolvimento Cientifico e Tecnologico and the Ministerio da Ciencia, Tecnologia e Inovacao, the Deutsche Forschungsgemeinschaft and the Collaborating Institutions in the Dark Energy Survey. The Collaborating Institutions are Argonne National Laboratory, the University of California at Santa Cruz, the University of Cambridge, Centro de Investigaciones Energeticas, Medioambientales y Tecnologicas-Madrid, the University of Chicago, University College London, the DES-Brazil Consortium, the University of Edinburgh, the Eidgenossische Technische Hochschule (ETH) Zurich, Fermi National Accelerator Laboratory, the University of Illinois at Urbana-Champaign, the Institut de Ciencies de l'Espai (IEEC/CSIC), the Institut de Fisica d'Altes Energies, Lawrence Berkeley National Laboratory, the Ludwig-Maximilians Universitat Munchen and the associated Excellence Cluster Universe, the University of Michigan, the National Optical Astronomy Observatory, the University of Nottingham, the Ohio State University, the University of Pennsylvania, the University of Portsmouth, SLAC National Accelerator Laboratory, Stanford University, the University of Sussex, and Texas A\&M University.

BASS is a key project of the Telescope Access Program (TAP), which has been funded by the National Astronomical Observatories of China, the Chinese Academy of Sciences (the Strategic Priority Research Program "The Emergence of Cosmological Structures" Grant \# XDB09000000), and the Special Fund for Astronomy from the Ministry of Finance. The BASS is also supported by the External Cooperation Program of Chinese Academy of Sciences (Grant \# 114A11KYSB20160057), and Chinese National Natural Science Foundation (Grant \# 11433005).

The Legacy Survey team makes use of data products from the Near-Earth Object Wide-field Infrared Survey Explorer (NEOWISE), which is a project of the Jet Propulsion Laboratory/California Institute of Technology. NEOWISE is funded by the National Aeronautics and Space Administration.

The Legacy Surveys imaging of the DESI footprint is supported by the Director, Office of Science, Office of High Energy Physics of the U.S. Department of Energy under Contract No. DE-AC02-05CH1123, by the National Energy Research Scientific Computing Center, a DOE Office of Science User Facility under the same contract; and by the U.S. National Science Foundation, Division of Astronomical Sciences under Contract No. AST-0950945 to NOAO.




\bibliographystyle{mnras}
\bibliography{project} 



\appendix

\section{Comparison with Z20 photometric redshifts}
\label{app:A}

The purpose of this appendix is to present a detailed comparison between our photometric redshifts and those of Z20 \citep{Zhou2020}, including the impact of the different photo-$z$ options on our cosmological results.

\begin{figure}
	\centering
	\includegraphics[width=0.47\textwidth]{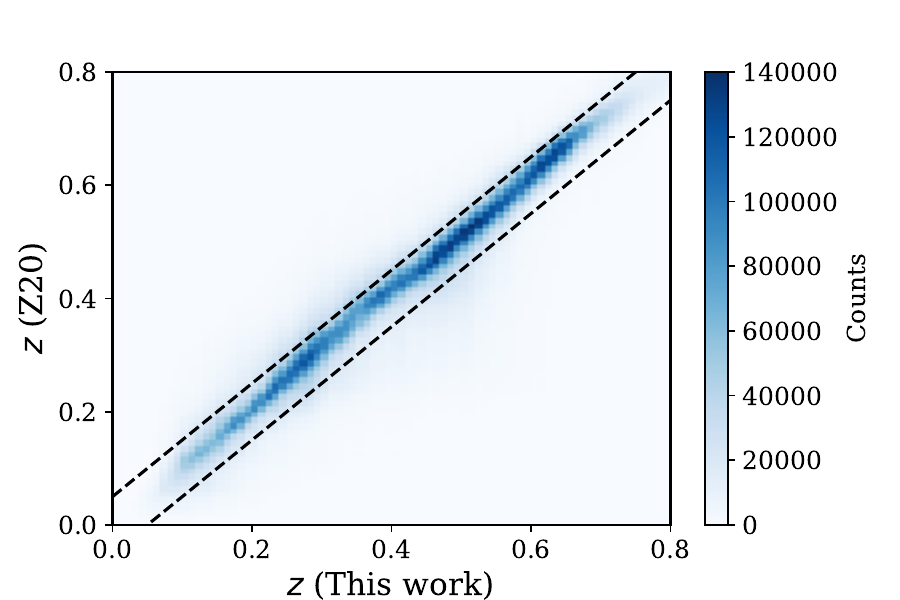}
	\caption{Photometric redshifts inferred from $g-r$, $r-z$, and $z-W_1$ colours, versus that from Z20. The dotted lines mark $|\Delta_z|=0.05$ interval. In the clipped sample, we only use objects inside the dashed line.}
	\label{fig:3dinferz_vs_zhouz}
\end{figure}

\begin{figure*}
	\centering
	\includegraphics[width=\textwidth]{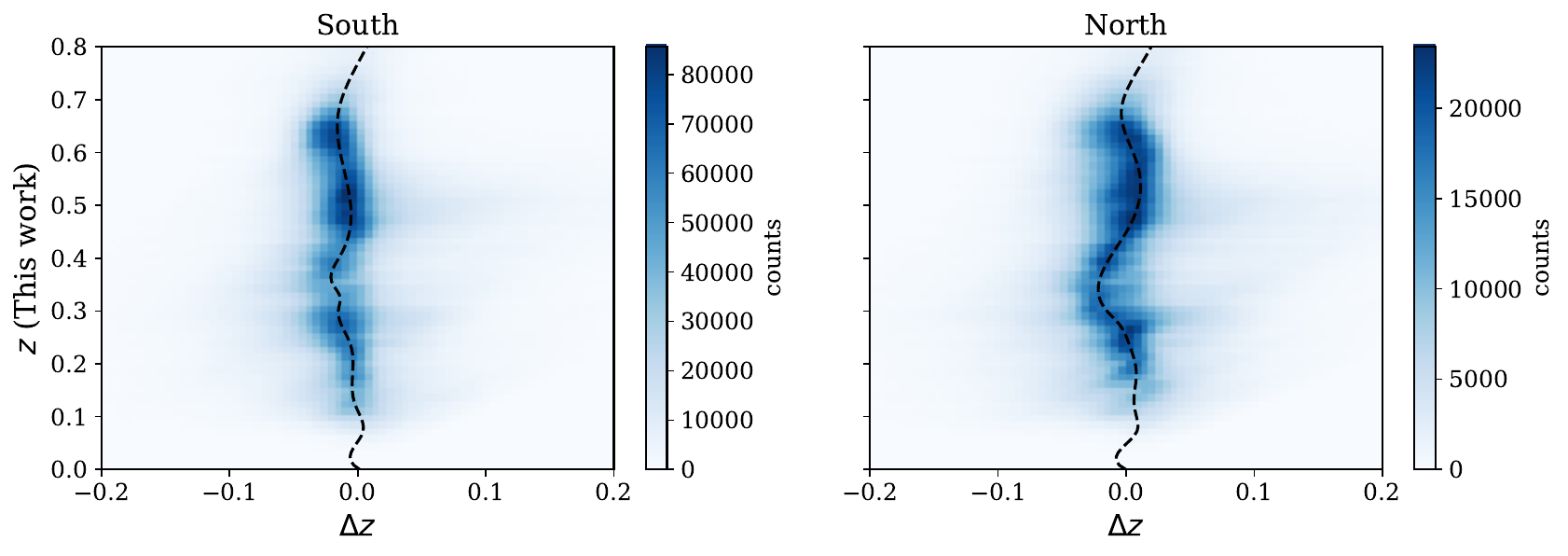}
	\caption{Photometric redshifts inferred from $g-r$, $r-z$, and $z-W_1$ colours, versus the difference from the Z20 estimates. The dotted lines show a spline fit to $\Delta z$ as a function of our photo-$z$, used for the offset correction.}
	\label{fig:deltaz_north_south}
\end{figure*}

\begin{figure*}
	\centering
	\includegraphics[width=\textwidth]{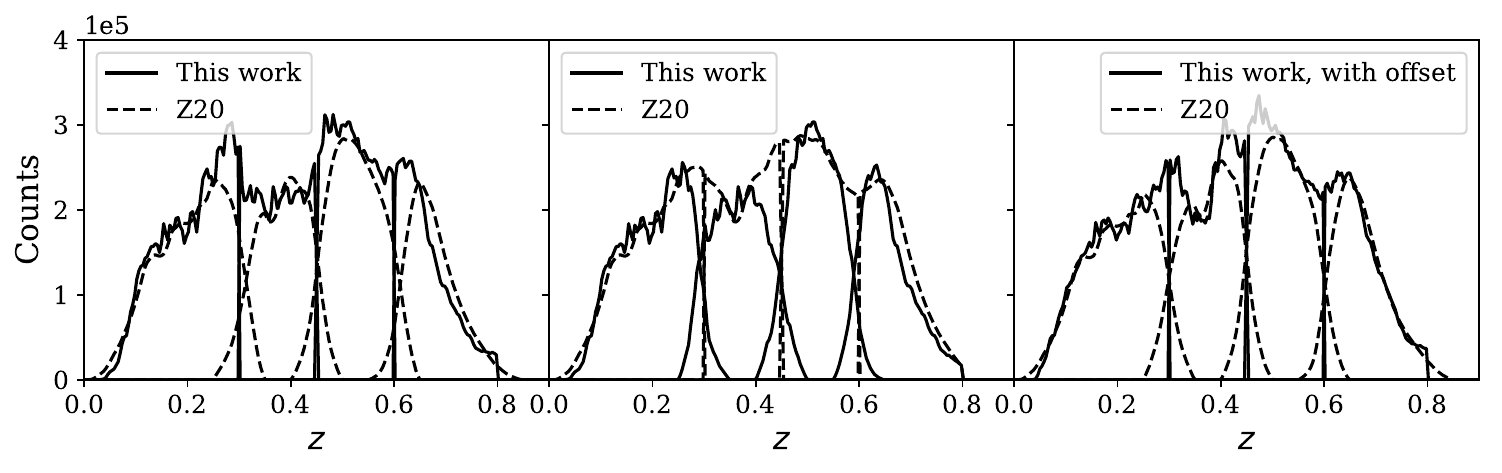}
	\caption{The raw redshift distribution binned using photo-$z$ obtained in this work (left), in Z20 (middle), and in this work with the correction for the offset (right), after a clipping of $|\Delta z|<0.05$. The solid line shows the distribution of photo-$z$ in this work, while the dashed line shows that from Z20.}
	\label{fig:nz_3dsets}
\end{figure*}

Fig.~\ref{fig:3dinferz_vs_zhouz} compares the two photo-$z$ catalogues in detail. The black dashed lines show the interval where the two photo-$z$ has a difference smaller than 0.05. We exclude objects outside the interval, cutting $23.4\%$ of the sample. Furthermore, there is a slight offset in the mean of the two samples, shown explicitly in the north and south part of the Legacy Survey in Fig.~\ref{fig:deltaz_north_south}. We fit this offset for the south and north part of the survey separately using a cubic spline. Then we further create an `offset' sample which has its redshifts corrected using the spline for $\Delta_z(z)$ to match with that of Z20. For this sample, the clipping of $|\Delta_z|=0.05$ is applied after correcting for the offset, cutting $22.5\%$ of the objects. Fig.~\ref{fig:nz_3dsets} compares the raw redshift distributions of this work and Z20 for the three samples. The left panel shows the sample using redshifts inferred from $g-r$, $r-z$, and $z-W_1$ colours, the middle panel shows that from Z20, and the right panel shows that from the offset sample. The two photo-$z$ distributions are close in all cases.

We find the photo-$z$ convolution function parameters, $(\sigma^{\rm spec},a^{\rm spec})$, for the Z20 samples using the same spectroscopic samples. We then follow the same procedures to find the best-fit $n(z)$. The parameters are summarized in Table~\ref{tab:zhou parameters}

\begin{table}
	\centering
	\caption{Photo-$z$ parameters for Z20.}
	\label{tab:zhou parameters}
\begin{tabular}{ l|l|l|l|l }
 \hline
 & bin 0 & bin 1 & bin 2 & bin 3\\ 
 \hline
 $\sigma^{\rm spec}$ & 0.0075 & 0.0128 & 0.0150 & 0.0248\\
 $a^{\rm spec}$ & 1.320 & 1.484 & 1.700 & 1.502\\
 $a^{\rm bfit}$ & 1.320 & 1.110 & 1.697 & 1.502\\
 $x_0^{\rm bfit}$ & 0.000 & 0.0003 & \llap{$-$}0.0002 & \llap{$-$}0.0001\\
 \hline
\end{tabular}
\end{table}


\bsp	
\label{lastpage}
\end{document}